\documentstyle[rawfonts,hip-artc_prepr,pslatex]{article}  
\volnumber{15}  \edyear{2002} \frompage{000} \topage{000}                %| 
\recrevdate{\today}                                                %| 
\def\rightmark{What is the Real $K$ Factor?} 
\def\leftmark{R. Vogt}
 
\title{What is the Real $K$ Factor?} 
\authors{
{R. Vogt}\\[2.812mm]
{\normalsize
Nuclear Science Division,
Lawrence Berkeley National Laboratory, \\ 
University of California, Berkeley, California 94720, USA\\
and\\
Physics Department, University of California, Davis, CA 95616, USA \\[0.2ex]
}}
 
\abstract{The theoretical $K$ factor, describing the difference 
between the leading and 
higher order cross sections, has no precise definition.  The definition is
sensitive to the order of the fit to the parton densities 
and the number of loops at 
which $\alpha_s$ is evaluated.  We describe alternate ways to define
the $K$ factor and show how the definition affects its magnitude and shape 
for examples of hadroproduction of $W^+$ bosons, Drell-Yan lepton pairs, and 
heavy quarks.  We discuss which definition is appropriate under
certain circumstances.}

 
\makeindex
\begin{document}

\maketitle

\section{Introduction}
\label{intro}

It has been clear for many years that only a leading order, LO, evaluation
of quantum chromodynamics, QCD, cross sections is inadequate to describe 
Drell-Yan, heavy quark, and jet 
production.  For example, at LO lepton pairs from the Drell-Yan process and 
heavy quark pairs are produced back-to-back with zero transverse momentum, 
$p_T$, of the pair.  Thus there is no nonzero Drell-Yan or $Q \overline Q$
pair $p_T$ at LO.  In addition, LO cross sections underestimate the measured 
total cross sections by up to a factor of two or more.

Next-to-leading order, NLO, evaluations of these cross sections removed many of
these inadequacies, especially concerning finite Drell-Yan or $Q \overline Q$ 
pair $p_T$
since light quark or gluon emission at NLO keeps the pair from being perfectly
back-to-back.  In addition, processes such as $gg \rightarrow g g^* \rightarrow
g Q \overline Q$ produce high $p_T$ $Q \overline Q$ pairs since the pair takes
the entire $p_T$ of the excited gluon.  The magnitude of the
Drell-Yan cross section at NLO is in much better agreement with the data.
Most modern parton distribution functions, PDFs, 
are evaluated at NLO and fewer new sets are
evaluated at LO.  However, only LO matrix elements are still employed for some
processes such as NRQCD quarkonia production, not yet fully calculated at NLO
\cite{Kramer}.  In addition, the LO matrix elements are inputs to event
generators like PYTHIA \cite{pythia}.  
Therefore, LO calculations are still of use, either for speed or ease of
calculation, if normalized properly relative to NLO.

Proper normalization is important to avoid over- or
underestimating the yields with a scaled-up LO calculation.  In addition,
the shapes of the resulting observable distributions may be different 
because the LO and NLO PDFs are not identical even
though they are fit to the same data.  The LO PDFs must be larger than 
the NLO PDFs in some regions since the NLO total cross sections are generally 
bigger and the LO PDFs must be larger to compensate.  
In some cases, certain contributions to NLO
evaluations are absent at LO.  For example, $\gamma^*$ and
vector boson production, denoted by $V$, proceeds only by $q \overline q$
annihilation, $q \overline q \rightarrow V$, at LO
while at NLO, $q \overline q \rightarrow Vg$ and the new process,
$q g \rightarrow V q$, with a gluon in the initial state are
also possible.  Thus, full NLO calculations, using NLO PDFs, 
two-loop calculations of $\alpha_s$ and NLO fragmentation functions, where
applicable, are preferable.  

The `so-called' $K$ factor is used to normalize the LO calculations. 
The theoretical $K$ factor is conventionally defined as the ratio between the 
NLO and LO cross sections.  This factor is generally assumed to be 
constant as a function of the relevant observables.  In heavy ion physics, 
the $K$ factor has often 
been assumed to be 2 without justification.  To be certain that
a LO calculation multiplied by a $K$ factor does not either over- or 
underestimate the NLO cross section, $K$ should be determined
according to a clearly defined prescription appropriate to the calculational
method.

There is more than one way to define
this factor.  Before NLO calculations were generally available, the LO
calculations were scaled up to the data by an arbitrary factor, the original
$K$ factor.  This is an experimental definition,
\begin{eqnarray}
K_{\rm exp} = \frac{\sigma_{\rm exp}}{\sigma_{\rm th}} \, \, .
\label{kexp}
\end{eqnarray}
If the NLO calculations do not agree with the data, this ratio of data to the
higher order calculation can also be defined, usually referred to now as 
`data/theory', see Ref.~\cite{matteo} for a recent usage.  
The $K$ factor is more often determined theoretically.
The usual theoretical $K$ factor is 
\begin{eqnarray}
K_{\rm th}^{(1)} = \frac{\sigma_{\rm NLO}}{\sigma_{\rm LO}}
\label{kth1}
\end{eqnarray}
where the superscript $(1)$ refers to the fact that only the first order
correction is included.  When further, higher order, corrections exist, as they
do, in part, for Drell-Yan and gauge boson production \cite{zwrefs} as well as
heavy quark production \cite{KLMV}, higher
order theoretical $K$ factors can also be defined, such as
\begin{eqnarray}
K_{\rm th}^{(2)} = \frac{\sigma_{\rm NNLO}}{\sigma_{\rm LO}} \, \, ,
\,\,\,\,\,\,\,\,\,\,\,\ 
K_{\rm th}^{(2')} = \frac{\sigma_{\rm NNLO}}{\sigma_{\rm NLO}} \, \, .
\label{kth2}
\end{eqnarray}
We do not go beyond $K_{\rm th}^{(1)}$ here.

There is some ambiguity in the definition of $K_{\rm th}^{(1)}$.  To 
demonstrate this, we discuss how the LO and NLO cross sections are calculated.
We first define 
\begin{eqnarray}
\sigma_{\rm LO} \equiv \sigma(\alpha_s^n) \, \, 
\end{eqnarray} 
where the exponent $n$ is defined as 
the power of $\alpha_s$ appropriate for the
tree-level partonic process.  We do not include the powers of the electroweak
coupling constant here for brevity and only concern ourselves with
hadroproduction.  Thus for Drell-Yan and gauge boson 
production, $n=0$ and $\sigma_{\rm LO} \equiv \sigma(1)$.  For direct
photon production via $q \overline q \rightarrow \gamma g$ and $qg \rightarrow 
\gamma g$, $n=1$ and $\sigma_{\rm LO} \equiv \sigma(\alpha_s)$.  Finally, for
heavy quarks produced through the $q \overline q \rightarrow Q \overline Q$ and
$gg \rightarrow Q \overline Q$ channels, $n=2$ and $\sigma_{\rm LO} \equiv
\sigma(\alpha_s^2)$. 
The total NLO cross section is the sum of the LO cross section and the
next-order correction, $\sigma(\alpha_s^{n+1})$,
with one additional power of $\alpha_s$, 
\begin{eqnarray}
\sigma_{\rm NLO} \equiv \sigma_{\rm LO} + \sigma(\alpha_s^{n+1}) = 
\sigma(\alpha_s^n) + \sigma(\alpha_s^{n+1}) \, \, 
\end{eqnarray}  
where we denote the total cross section at NLO as $\sigma_{\rm NLO}$.

The next order correction for Drell-Yan and gauge boson production, along with
virtual corrections to $q \overline q \rightarrow V$, are the new processes
$q \overline q
\rightarrow V g$ and $qg \rightarrow Vq$.  They are first order in $\alpha_s$,
so that $\sigma_{\rm NLO}^{\rm DY} = \sigma(1)
+\sigma(\alpha_s)$.  The NLO correction to heavy quark hadroproduction is
$\sigma(\alpha_s^3)$ so that $\sigma_{\rm NLO}^{Q \overline Q} = 
\sigma(\alpha_s^2) +
\sigma(\alpha_s^3)$.  The NLO contribution, $\sigma(\alpha_s^{n+1})$, 
is calculated with $\alpha_s$
evaluated to two loops and with NLO PDFs.
However, there is some ambiguity in how to calculate $\sigma_{\rm LO}$.  In
principle, at each order of $\alpha_s$, there should be an appropriate PDF, LO
for $\sigma(\alpha_s^n)$, NLO for $\sigma(\alpha_s^{n+1})$, NNLO for
$\sigma(\alpha_s^{n+2}) \cdots$ with $\alpha_s$ evaluated to one-, two-, and
three-loop accuracy respectively.  Although possible, at least to NLO, this is
usually not done.  The NLO PDFs and two-loop $\alpha_s$ are typically used as
defaults for all orders of the cross section.  In our calculations, we will
compare the magnitude and shape of the cross sections with $\sigma_{\rm LO}$
evaluated employing LO PDFs and one-loop $\alpha_s$, a full LO cross section,
and calculations of $\sigma_{\rm LO}$ employing NLO PDFs and two-loop
$\alpha_s$.  We also investigate the effect of these two definitions of
$\sigma_{\rm LO}$ on the theoretical $K$ factor for the processes considered.

Thus if $\sigma_{\rm LO}$ is fully LO, we define
\begin{eqnarray}
\sigma_{\rm LO(1)} & \equiv & \sigma_1(\alpha_s^n) \, \, , \label{sfullLO} \\
\sigma_{\rm NLO(1)} & \equiv & \sigma_1(\alpha_s^n) + \sigma(\alpha_s^{n+1}) \,
\,  \label{sLOpNLO}
\end{eqnarray}
where the subscript `1' refers to the LO PDFs and one-loop $\alpha_s$.
If, instead the LO cross section is evaluated with the same PDFs and $\alpha_s$
as the NLO correction, we have
\begin{eqnarray}
\sigma_{\rm LO(2)} & \equiv & \sigma_2(\alpha_s^n) \, \, , \label{sNLOLO} \\
\sigma_{\rm NLO(2)} & \equiv & \sigma_2(\alpha_s^n) + \sigma(\alpha_s^{n+1}) \,
\,  \label{sfullNLO}
\end{eqnarray}
where now the subscript `2' refers to the NLO PDFs and two-loop $\alpha_s$.
Recall that the 
NLO correction, $\sigma(\alpha_s^{n+1})$, is always calculated with NLO 
PDFs and the two-loop evaluation of $\alpha_s$.

The total NLO cross section is usually calculated
with NLO PDFs and two-loop $\alpha_s$ at each order, 
as in Eq.~(\ref{sfullNLO}), to determine the size of
the next-order correction independent of the shape of the PDF and difference in
magnitude of $\alpha_s$.  Thus, the most typical way to evaluate the NLO
theoretical $K$ factor in Eq.~(\ref{kth1}) is with Eqs.~(\ref{sNLOLO}) and
(\ref{sfullNLO}), 
\begin{eqnarray}
K_{\rm th, \, 0}^{(1)} \equiv \frac{\sigma_2(\alpha_s^n) +
\sigma(\alpha_s^{n+1})}{\sigma_2(\alpha_s^n)} = \frac{\sigma_{\rm
NLO(2)}}{\sigma_{\rm LO(2)}} \, \, .
\label{kthdef0}
\end{eqnarray}
We have labeled this as the $0^{\rm th}$ definition because it is most often
used by theorists, see Ref.~\cite{zwrefs} for an example.
It is perhaps more correct to define $K_{\rm th}^{(1)}$ based on
$\sigma_{\rm LO(1)}$, as in Eqs.~(\ref{sfullLO}) and (\ref{sLOpNLO}),
\begin{eqnarray}
K_{\rm th, \, 1}^{(1)} \equiv \frac{\sigma_1(\alpha_s^n) +
\sigma(\alpha_s^{n+1})}{\sigma_1(\alpha_s^n)} = \frac{\sigma_{\rm
NLO(1)}}{\sigma_{\rm LO(1)}} \, \, .
\label{kthdef1}
\end{eqnarray}
Now the subscript `1' on $K_{\rm th, \, 1}^{(1)}$ 
refers to the order of the PDFs and $\alpha_s$ at which 
$\sigma_{\rm LO}$ is evaluated.
However, if $\sigma_{\rm NLO(2)}$, is known 
but the LO calculation, $\sigma_{\rm LO(1)}$, is most
convenient, as in an event generator, 
it might be more advantageous to define the $K$ factor as
\begin{eqnarray}
K_{\rm th, \, 2}^{(1)} \equiv \frac{\sigma_2(\alpha_s^n) +
\sigma(\alpha_s^{n+1})}{\sigma_1(\alpha_s^n)} = \frac{\sigma_{\rm
NLO(2)}}{\sigma_{\rm LO(1)}} \, \, .
\label{kthdef2}
\end{eqnarray}
This last definition maximally mixes 
the PDF and $\alpha_s$ evaluations.  Thus $K_{\rm th, \, 2}^{(1)}$ 
is most dependent on the difference in the shapes of the LO and NLO 
PDF fits.  Note that, in this case only, the $K$ factor cannot be written as 
$K\sim 1 + {\cal O}(\alpha_s)$.
The expressions above are written in terms of the total cross
sections but are also equally valid for the differential distributions.

We calculate the differential $K_{\rm th, \, 0}^{(1)}$,  
$K_{\rm th, \, 1}^{(1)}$,  and $K_{\rm th, \, 2}^{(1)}$ with $W^+$, Drell-Yan
and $Q \overline Q$ production as specific examples.  
In the case of vector bosons, $K_{\rm th}^{(1)}$ should
most strongly reflect the difference between the LO and NLO PDF evaluations
while $Q \overline Q$ production is quite sensitive to the order of the
$\alpha_s$ evaluation.
See Ref.~\cite{failevai}
for a study of $K_{\rm th, \, 0}^{(1)}$ in jet production.
In our calculations, we use the MRST HO (central gluon)
\cite{mrst} PDFs in the ${\overline
{\rm MS}}$ scheme for $\sigma(\alpha_s^{n+1})$ and $\sigma_{\rm LO(2)}$ 
and the MRST LO (central gluon)
\cite{mrstlo} PDFs for $\sigma_{\rm LO(1)}$ unless otherwise specified.

\section{$W^+$ Production in $pp$ Interactions}
\label{wp}

We begin our discussion of the $K$ factors
with the $W^+$ rapidity distributions.  This process is independent of the 
order of $\alpha_s$ since at NLO $\alpha_s$ is always evaluated at two loops.
It is also the least differential of all the processes considered.  
We choose the $W^+$ of the three gauge bosons because
its rapidity distribution is a strong function of the shape of the quark
PDFs at LO and NLO.  We show our results at the top $pp$ energy of 
the LHC, 14 TeV.  The dependence of $K_{\rm th}^{(1)}$ on rapidity is
similar for all gauge bosons so that one case is sufficient for illustration. 
See Ref.~\cite{RVwz}
for a comparison of $K_{\rm th, \, 0}^{(1)}$
for all three gauge bosons.  
We do not discuss the $p_T$ dependence of either
$W^+$ or Drell-Yan production because the $p_T$ of both are zero at LO.
Thus $K_{\rm th}^{(1)}(p_T)$ cannot be defined for $W^+$ production, 
only $K_{\rm th}^{(2)}(p_T)$. 

The next-to-leading order, NLO, cross section for production of a
vector boson, $V$, with mass $m_V$ at scale $Q^2$ in a $pp$ interaction is
\begin{eqnarray} \frac{d\sigma^V_{pp}}{dy} & = & 
H_{ij}^V \int dx_1\, dx_2 \,dx \, \delta
\bigg(\frac{m_V^2}{S} - x x_1 x_2 \bigg) 
\delta \bigg( y - \frac{1}{2} \ln \bigg( \frac{x_1}{x_2} \bigg) \bigg) 
\label{sigmajpsi} \\
&   & \mbox{} \times
\bigg\{ \sum_{q_i,q_j \in Q,\overline Q} C^{\rm ii}(q_i, \overline q_j)
\Delta_{q \overline q}(x,Q^2) f_{q_i}^p(x_1,Q^2) f_{\overline 
q_j}^p(x_2,Q^2) \nonumber \\
&   & \mbox{} +
\sum_{q_i,q_k \in Q, \overline Q} 
C^{\rm if}(q_i, q_k) \Delta_{qg}(x,Q^2) \bigg[
f_{q_i}^p(x_1,Q^2) f_g^p(x_2,Q^2) \nonumber \\
&   & \mbox{} + f_g^p(x_1,Q^2) 
f_{q_j}^p(x_2,Q^2) \bigg] \bigg\} \, \,  , \nonumber
\end{eqnarray}
where $m_V$ is the boson mass, $S$ is the center of mass energy squared,
$H_{ij}^V$ is proportional to the LO 
partonic cross section, $q_i q_j \rightarrow V$, 
and the sum over $q_i$ runs from $u$
to $c$.  At this energy $Q^2 \gg m_c^2$ so that the $c$ quark contribution 
cannot be neglected.  The matrices $C^{\rm ii}$ and
$C^{\rm if}$ contain information on the coupling of the various quark flavors
to boson $V$.  The parton densities in the proton are given by
$f_{q_i}^p(x,Q^2)$ and 
are evaluated at momentum fraction $x$ and scale $Q^2$.  For a vector boson
without a fixed mass, as in the Drell-Yan process, the
mass distribution can also be calculated by adding a $dM$ in the denominator of
the left-hand side of Eq.~(\ref{sigmajpsi}) and the delta function $\delta(M -
m_V)$ on the right-hand side.

The prefactors $H_{ij}^V$ are rather simple \cite{zwrefs}.  In the case of
$W^+$ production,
\begin{eqnarray}
H_{ij}^{W^\pm} = \frac{2\pi}{3} \frac{G_F}{\sqrt{2}} \frac{m_W^2}{S} \, \, ,
\label{wpmpart} 
\end{eqnarray}
where $G_F = 1.16639 \times 10^{-5}$ GeV$^2$ and $m_W = 80.41$ GeV.  
For virtual photon production via the Drell-Yan process,
there are three contributions to $H_{ii}^M$: virtual photon exchange, 
$Z^0$ exchange, and $\gamma^*-Z^0$ interference,
\begin{eqnarray}
H_{ii}^M & = & H_{ii}^{\gamma^*} + H_{ii}^{\gamma^*-Z^0} + H_{ii}^{Z^0} : 
\label{dypart} \\
H_{ii}^{\gamma^*} & = & \frac{4 \pi \alpha^2}{9 M^2 S} |e_i|^2 \nonumber \\
H_{ii}^{\gamma^*-Z^0} & = & \frac{\alpha}{9}\frac{G_F}{\sqrt{2}} 
\frac{m_Z^2}{S}
\frac{(1 - 4\sin^2\theta_W)(M^2 - m_Z^2)}{(M^2 - m_Z^2)^2 + m_Z^2 \Gamma_Z^2}
|e_i|(1 - 4|e_i|\sin^2\theta_W) \nonumber \\
H_{ij}^{Z^0} & = & \frac{1}{3} \frac{G_F}{\sqrt{2}} \frac{M^2}{S}
\frac{m_Z \Gamma_{Z \rightarrow l^+ l^-}}{(M^2 - m_Z^2)^2 + m_Z^2 \Gamma_Z^2}
(1 + (1 - 4|e_i| \sin^2\theta_W)^2)
\nonumber 
\end{eqnarray}
where $\sin^2 \theta_W = 1 - m_W^2/m_Z^2$, $m_Z = 91.187$ GeV, $\Gamma_Z = 
2.495$ GeV, and $\Gamma_{Z \rightarrow l^+ l^-} = 3.367$\%.
Note that here we use $i=j$ since $i$ and $j$ must be the same flavor for
neutral gauge boson production.  Now $H_{ii}^M$ depends on the pair 
mass and goes inside the mass integral.

The functions $\Delta_{ij}(x,Q^2)$ in Eq.~(\ref{sigmajpsi}) are 
universal for all $V$ \cite{zwrefs}.  We work in the
${\overline {\rm MS}}$ scheme.  The NLO correction to the $q \overline q$
channel includes the contributions from soft and virtual gluons as well as hard
gluons from the process $q \overline q \rightarrow V g$.  We have, up to NLO
\cite{zwrefs},
\begin{eqnarray} 
\Delta_{q \overline q}(x,Q^2) & = & \Delta_{q \overline q}^{\rm
LO}(x,Q^2) + \Delta_{q \overline q}^{\rm NLO}(x,Q^2)
\label{delqqbar} \\  
\Delta_{q \overline q}^{\rm LO}(x,Q^2) & = & \delta(1-x) \label{delqqbarLO} \\
\Delta_{q \overline q}^{\rm NLO}(x,Q^2) & = & \frac{\alpha_s(Q^2)}{3\pi} 
\bigg\{ -4(1 + x) \ln\bigg(\frac{Q^2}{m_V^2} \bigg)
-8(1+x) \ln(1-x)  \label{delqqbarNLO} \\
&   & \mbox{} - 4 \frac{1+x^2}{1-x} \ln x + \delta(1-x) 
\bigg[ 6 \ln \bigg(\frac{Q^2}{m_V^2} \bigg) + 8
\zeta(2) - 16 \bigg] \nonumber \\ &  & \mbox{} + 
8 \bigg[ \frac{1}{1-x} \bigg]_+  
\ln\bigg(\frac{Q^2}{m_V^2} \bigg)
+ 16 \bigg[ \frac{\ln(1-x)}{1-x} \bigg]_+ \bigg\} \, \, . \nonumber
\end{eqnarray}
The LO contribution is a delta function.  The NLO contribution is
proportional to $\alpha_s$, calculated to two loops
with $n_f = 5$ active flavors.  The first three terms of $\Delta_{q \overline
q}^{\rm NLO}(x,Q^2)$ 
are the real contributions from $q \overline q \rightarrow Vg$
while the last three terms are the soft and virtual
gluon contributions from the $q \overline q$ vertex correction.  
The general integral of the `plus' functions from soft gluon emission in 
Eq.~(\ref{delqqbar}) is 
\cite{hsv}
\begin{eqnarray}
\int_a^1 dx f(x) \bigg[ \frac{\ln^i (1-x)}{1-x} \bigg]_+ = \int_a^1 dx 
\frac{f(x)
- f(1)}{1-x} \ln^i(1-x) + \frac{f(1)}{i+1} \ln^{i+1}(1-a) \, \, .
\label{plus}
\end{eqnarray}
The quark-gluon contribution only appears at NLO through
the real correction $qg \rightarrow qV$.  At this order \cite{zwrefs}, 
\begin{eqnarray}
\Delta_{qg}(x,Q^2) & = & \Delta_{qg}^{\rm NLO}(x,Q^2)  \label{delqg} \\
 & = & \frac{\alpha_s(Q^2)}{8\pi} 
\bigg\{ 2(1 + 2x^2 - 2x)\ln \bigg(
\frac{(1-x)^2Q^2}{xm_V^2} \bigg) + 1 - 7x^2 + 6x \bigg\} \, \, . \nonumber
\end{eqnarray}
Note that, since at NLO, this contribution is only real, no delta functions or
plus distributions appear.  Using the delta functions in Eq.~(\ref{sigmajpsi}) 
we find  $x_{1,2} = \sqrt{Q^2/xS}
\exp( \pm y)$ at NLO.  For $\Delta_{q \overline q}^{\rm LO}(x,Q^2)$ and the
terms in $\Delta_{q \overline q}^{\rm NLO}(x,Q^2)$ contributing to the vertex
correction in Eq.~(\ref{delqqbarNLO}), $x_{1,2}' = 
\sqrt{Q^2/S} \exp( \pm y)$.  
If we take $Q^2 = m_V^2$, all terms proportional to 
$\ln(Q^2/m_V^2)$ drop out.  

We now define the coupling matrices, $C^{\rm ii}$ and $C^{\rm if}$, 
in Eq.~(\ref{sigmajpsi}).  The
superscripts represent the initial (i) and final (f) state quarks or antiquarks
while the arguments indicate
the orientation of the quark line to which the boson is coupled \cite{zwrefs}.
The $W^+$ couplings are elements of the CKM matrix.
They are nonzero for $C^{\rm ii}(q_k,\overline q_l)$ if $e_k + e_l = \pm 1$
and for $C^{\rm if}(q_k,q_l)$ if $e_k = \pm 1 + e_l$.  In both cases, they take
the values $| V_{q_k q_l} |^2$.  Following Ref.~\cite{zwrefs},
we take $V_{ud} = \cos \theta_C \approx V_{cs}$ and $V_{us} =
\sin \theta_C \approx -V_{cd}$ with $\sin \theta_C \approx 0.22$.  

The $W^+$ $q \overline q$ convolution in a $pp$ interaction is
\begin{eqnarray}
\lefteqn{ \sum_{q_i,q_j \in Q, \overline Q} 
f_{q_i}^p(x_1,Q^2) f_{\overline 
q_j}^p(x_2,Q^2) C^{\rm ii}(q_i, \overline q_j)}
\label{dswpdyqqb} \\ & & \mbox{} =
\cos^2 \theta_C
\left(f_u^p(x_1,Q^2) f_{\overline d}^p(x_2,Q^2) + 
f_{\overline s}^p(x_1,Q^2) f_c^p(x_2,Q^2) \right)
\nonumber \\ & & \mbox{} + \sin^2 \theta_C
\left(f_u^p(x_1,Q^2)  f_{\overline s}^p (x_2,Q^2) +
f_{\overline d}^p(x_1,Q^2) f_c^p (x_2,Q^2) \right) + 
[ x_1 \leftrightarrow x_2] \, \, . \nonumber
\end{eqnarray}
For the $qg$ channel, we have
\begin{eqnarray}
\lefteqn{ \sum_{q_i,q_k \in Q \overline Q} \bigg(
f_{q_i}^p(x_1,Q^2) f_g^p(x_2,Q^2) + 
[ x_1 \leftrightarrow x_2 ] \bigg)
C^{\rm if}(q_i, q_k)}
\label{dswpdyqg} \\ & & \mbox{}  = f_g^p(x_2,Q^2) \left(
f_u^p(x_1,Q^2) + f_{\overline d}^p(x_2,Q^2) + f_{\overline
s}^p (x_1,Q^2) + f_c^p(x_1,Q^2) \right) \nonumber \\
&  & \mbox{}
+ [ x_1 \leftrightarrow x_2 ] \, \, .
\nonumber
\end{eqnarray}
The couplings do not enter explicitly in the $qg$ convolution
because each
distribution is multiplied by $(\cos^2 \theta_C + \sin^2 \theta_C)$.  

\begin{figure}[htb]
\setlength{\epsfxsize=0.75\textwidth}
\setlength{\epsfysize=0.4\textheight}
\centerline{\epsffile{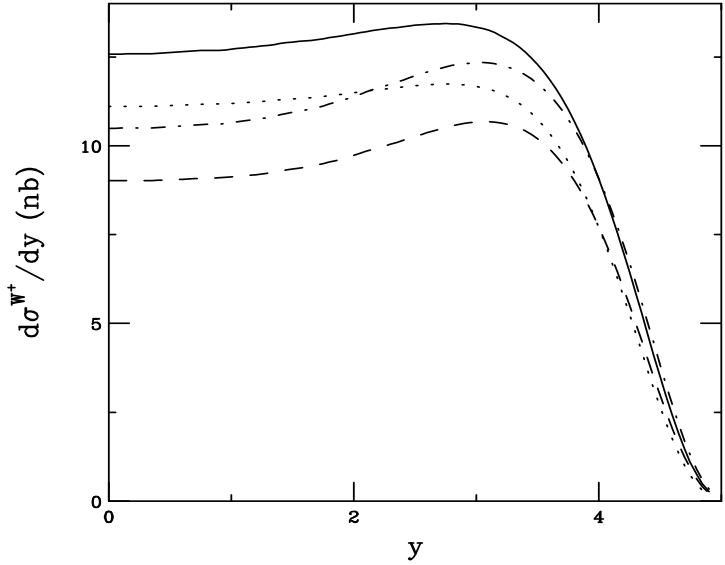}}
\caption{The $W^+$ rapidity distribution in $pp$ collisions at 14 TeV 
evaluated at $Q = m_{W^+}$.  The solid curve is $\sigma_{\rm NLO(2)}$, 
the dashed curve is $\sigma_{\rm LO(1)}$, the dot-dashed curve is 
$\sigma_{\rm NLO(1)}$ and the dotted curve is $\sigma_{\rm LO(2)}$.  
}
\label{wpdist}
\end{figure}

In Fig.~\ref{wpdist} we show the $W^+$ rapidity distributions calculated with
Eqs.~(\ref{sfullLO})-(\ref{sfullNLO}).
The `standard' NLO calculation, $\sigma_{\rm NLO(2)}$, 
is shown in the solid
curve.  The accompanying LO result, $\sigma_{\rm LO(2)}$, is given 
in the dotted
curve.  The two curves are nearly parallel to each other over most of the
available rapidity space.  Both increase slowly with rapidity until $y \sim 3$
where the turnover point is reached.  The NLO calculation increases slightly 
faster than the LO, most likely due to the $qg$ contribution, $f_u^p(x_1) 
f_g^p(x_2)$, since the gluon distribution 
increases with decreasing $x$ when $x$
is small.  The increase with $y$ is due to the $f_u^p(x_1) f_{\overline d}^p
(x_2)$ component of the $q \overline q$ annihilation cross section.  The 
valence $u$ distribution in the proton, $f_{u_V}^p(x_1)$, is increasing until
$x_1 \sim 0.126$, corresponding to the peak of $f_{u_V}^p(x_1)$ at
$y \sim 3$.  Likewise $f_{\overline d}^p(x_2)$  is rising
with decreasing $x_2$.  Only when $f_{u_V}^p(x_1)$ is falling with
growing $x_1$, beyond the peak, does the cross section begin to decrease.  

The same effect is seen, 
albeit more strongly, for $\sigma_{\rm NLO(1)}$ and $\sigma_{\rm LO(1)}$.  
We see that $\sigma_{\rm LO(2)}$ increases 6\% for $0<y<3$ 
while $\sigma_{\rm LO(1)}$ 
increases by 18\% over the same interval.  The difference
arises primarily because $f_{\overline
d}^p(x)$ increases faster with decreasing $x$ at LO than at NLO.
Note also that $\sigma_{\rm LO(2)}$ is about 20\% larger than $\sigma_{\rm
LO(1)}$ at $y=0$.
This difference is due to the fact that the NLO PDFs are larger in the region 
around $x \sim 5.6 \times 10^{-3}$.  As $y$ increases, the LO valence and sea
distributions become closer to the NLO distributions, so that, in the tail,
$\sigma_{\rm LO(1)} > \sigma_{\rm LO(2)}$.  
This trend is also reflected in the corresponding
NLO distributions.

\begin{figure}[htb]
\setlength{\epsfxsize=0.75\textwidth}
\setlength{\epsfysize=0.4\textheight}
\centerline{\epsffile{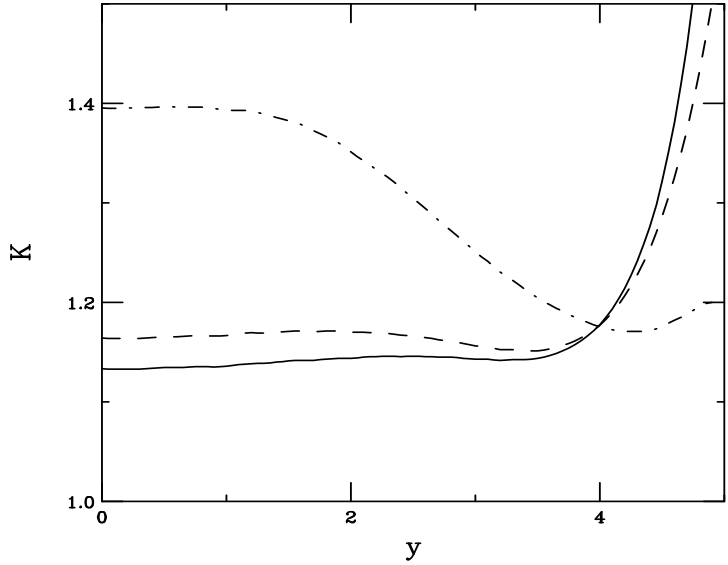}}
\caption{The three $K$ factors for $W^+$ production.  The solid curve is
$K_{\rm th, \, 0}^{(1)}$, the dashed is $K_{\rm th, \, 1}^{(1)}$, and the 
dot-dashed is $K_{\rm th, \, 2}^{(1)}$.
}
\label{wpkfac}
\end{figure}

The resulting $K$ factors are shown in Fig.~\ref{wpkfac}.  
The `standard' $K$
factor, $K_{\rm th, \, 0}^{(1)}$, is the smallest, $\sim 1.13$, for $0 < y <
3.5$ where it begins to increase.  The rise is due to the greater importance 
of the $f_q^p(x_1) f_g^p(x_2)$ contribution as $x_2 \rightarrow 0$.  The factor
$K_{\rm th, \, 1}^{(1)}$ is only
slightly higher, $\sim 1.16$, and also rather flat over most of the rapidity
range.  The largest $K$ factor is $K_{\rm th, \, 2}^{(1)}$, as might be
expected from inspection of the curves in Fig.~\ref{wpdist}, due to the
greater difference between $\sigma_{\rm NLO(2)}$ and
$\sigma_{\rm LO(1)}$.  The decrease for $y>2$ is due to the
steeper rise of the LO rapidity distribution as the peak of $f_{u_V}^p(x)$ 
is approached.  The minimum of $K_{\rm th, \, 2}^{(1)}$ is reached when
$\sigma_{\rm NLO(2)}$ drops below $\sigma_{\rm NLO(1)}$.

We have also checked the dependence of $K$ on scale and PDF.  The scale 
dependence enters not only in the PDFs and $\alpha_s$ but also in the 
logarithms $\ln (Q^2/m_V^2)$ in Eqs.~(\ref{delqqbar}) and (\ref{delqg}).
When $Q^2 = m_V^2$, as in our calculations so far, the three scale dependent
terms in $\Delta_{q \overline q}^{\rm NLO}(x,Q^2)$ drop out and  
$\Delta_{qg}^{\rm NLO}(x,Q^2)$
only depends on $\ln ((1-x)^2/x)$.  If $Q^2 > m_V^2$, the logs are
positive, enhancing $\sigma(\alpha_s)$, but if $Q^2 < m_V^2$,
the scale-dependent logs change sign, decreasing $\sigma(\alpha_s)$.
Scale evolution increases the low $x$ density of the sea
quarks and gluons while depleting the high $x$ component and
reducing the valence distributions.  The $x$ values do not change when $Q^2$
is varied.  Thus the higher scales also tend to enhance the cross sections.
At these large values of $Q^2$, $\alpha_s$ does not vary strongly so that 
increasing or decreasing $Q^2$ by a factor of 16 is not a large effect.
The total cross sections are larger when $Q^2 = 4m_V^2$ and smaller when $Q^2 
= m_V^2/4$, opposite the scale dependence at lower energies, due to the larger 
parton densities at low $x$ and high $Q^2$.  The same trend was also seen in 
Ref.~\cite{zwrefs}.
We find that all three defintions of the $K$ factor exhibit the same $Q^2$
dependence.  Increasing $Q^2$ gives a 7\% larger $K$ factor while decreasing 
$Q^2$ reduces the $K$ factors by $\sim 10$\% for the MRST set.

The variation of the results with PDF is also significant.  The MRST and
CTEQ \cite{cteq5} results are similar but the CTEQ cross sections are a few
percent higher.  In this case, the $\sigma_{\rm LO(2)}$ and $\sigma_{\rm
NLO(2)}$ distributions are not as broad as $\sigma_{\rm LO(1)}$ and
$\sigma_{\rm NLO(1)}$.  The peak of $\sigma_{\rm LO(2)}$ is at $y<3$ while
$\sigma_{\rm LO(1)}$ peaks at $y \sim 3.2$.  Because of this
difference, $K_{\rm th, \, 2}^{(1)}$ is not independent of rapidity anywhere 
but decreases almost to unity where $\sigma_{\rm LO(1)} \sim
\sigma_{\rm NLO(2)}$.  When the GRV 98 
distributions \cite{grv98} are used, the rise with rapidity
is much stronger than that in Fig.~\ref{wpdist}.  However, now the differences
between $\sigma_{\rm LO(1)}$ and $\sigma_{\rm LO(2)}$ 
are rather small, $\sigma_{\rm LO(1)}$ peaks about 0.25 units of rapidity 
higher than
$\sigma_{\rm LO(2)}$ and with a slightly larger value.  In this case, the
$K$ factors are all smaller than for MRST, $K_{\rm th, \, 0}^{(1)} \sim
K_{\rm th, \, 1}^{(1)} \sim 1.09$.  Now
even $K_{\rm th, \, 2}^{(1)}$ is only a few percent different than the other
two definitions over two units of rapidity.

\section{Drell-Yan Production in Pb+Pb Interactions}
\label{dy}

We now discuss dilepton production in the Drell-Yan process where $\alpha_s$
also does not enter until NLO, as in the previous section, but the LO cross
sections are more differential, depending on pair mass, $M$, and rapidity.
Since, at collider energies,
the Drell-Yan contribution to the dilepton continuum is small compared to heavy
quark decays and will probably not be cleanly separated, we discuss the
Drell-Yan yield at the CERN SPS.  The Drell-Yan cross section is of interest
here because it is the only source of lepton pairs in the mass region above
the $\psi'$ for $\sqrt{S} = 17.3$ GeV.  
As such, it has been used to normalize the $J/\psi$ yield and
determine its suppression as a function of transverse energy, $E_T$.  (See
Ref.~\cite{RVrev} for more details.)  The NA50 collaboration \cite{na50}
measures the
dilepton continuum above 4 GeV in Pb+Pb collisions
and fits the experimental $K$ factor to the dilepton yield in the mass
range $4 < M < 9$ GeV 
by a LO Drell-Yan calculation with the MRSA \cite{mrsa} NLO PDFs.
The Drell-Yan contribution below the $J/\psi$ peak, in the range $2.9 < M <
4.5$ GeV for NA50, 
is then calculated at LO and multiplied by this same experimental
$K$ factor, assuming that it is independent of mass and rapidity in the 
interval $0<y<1$.  It is appropriate to check this assumption.

The Drell-Yan cross section as a function of mass and rapidity is also given by
Eq.~(\ref{sigmajpsi}) with the delta function added for the mass, as explained
earlier. We do not include any nuclear effects on the PDFs.
The convolution of the prefactors $H_{ii}^M$, Eq.~(\ref{dypart}), with the
parton distribution functions in a nuclear collision
in the $q \overline q$ channel is:
\begin{eqnarray}
\lefteqn{ \sum_{q_i \in Q} 
H_{ii}^M  \bigg( f_{q_i}^N(x_1,Q^2) f_{\overline 
q_i}^N(x_2,Q^2) + f_{\overline q_i}^N(x_1,Q^2) f_{q_i}^N(x_2,Q^2) \bigg) }
\label{dszdyqqb} \\ & & \mbox{}  =
H_{uu}^M \left(
\left\{ Z_A f_u^p(x_1,Q^2)+ N_A f_u^n(x_1,Q^2) \right\}
\left\{ Z_B f_{\overline u}^p(x_2,Q^2) + N_B 
f_{\overline u}^n(x_2,Q^2)\right\} \right.
\nonumber \\ & & \left. \mbox{} + 2 AB
f_c^p(x_1,Q^2) f_{\overline c}^p(x_2,Q^2) \right)
\nonumber \\ & & \mbox{} +
H_{dd}^M \left(
\left\{ Z_A f_d^p(x_1,Q^2)+ N_A f_d^n(x_1,Q^2) \right\} 
%\nonumber \\ & & \mbox{} \times \left. \left.\left.
\left\{ Z_B f_{\overline d}^p(x_2,Q^2) + N_B 
f_{\overline d}^n(x_2,Q^2) \right\}  \right. 
\nonumber \\ & & \left. \mbox{} +
2 AB f_s^p(x_1,Q^2) f_{\overline s}^p(x_2,Q^2) \right) 
+ [ x_1 \leftrightarrow x_2 , A \leftrightarrow B ] \, \, .
\nonumber
\end{eqnarray}
We let $H_{uu}^M$ represent all charge $+2/3$ quarks and $H_{dd}^M$ all charge
$-1/3$ quarks.
In the $qg$ channel the convolution is
\begin{eqnarray}
\lefteqn{ \sum_{q_i \in Q, \overline Q} H_{ii}^M \bigg(
f_{q_i}^N(x_1,Q^2) f_g^N(x_2,Q^2) + f_g^N(x_1,Q^2) f_{q_i}^N(x_2,Q^2) 
\bigg)} \label{dszdyqg} \\ & & \mbox{}  = B f_g^p(x_2,Q^2) \bigg\{
H_{uu}^M \left(
\left\{ Z_A f_u^p(x_1,Q^2)+ N_A f_u^n(x_1,Q^2) \right\} 
\right. \nonumber \\ & & \mbox{} \left. + \left\{ Z_B 
f_{\overline u}^p(x_2,Q^2) + N_B 
f_{\overline u}^n(x_2,Q^2)\right\} + 2 A f_c^p(x_1,Q^2) \right)
\nonumber \\ & & \mbox{} +
H_{dd}^M \left(
\left\{ Z_A f_d^p(x_1,Q^2)+ N_A f_d^n(x_1,Q^2) \right\}  
\right. \nonumber \\ & & \mbox{} \left. 
+ \left\{ Z_B f_{\overline d}^p(x_2,Q^2) + N_B 
f_{\overline d}^n(x_2,Q^2) \right\} + 2 A f_s^p(x_1,Q^2) \right) \bigg\} 
\nonumber \\ 
& & \mbox{} + [ x_1 \leftrightarrow x_2, A \leftrightarrow B ] \, \, .
\nonumber
\end{eqnarray}
We have taken the isospin of the nuclei into account, denoting the proton and
neutron numbers in a nucleus with mass $A$ by $Z_A$ and $N_A$ respectively.

\begin{figure}[htb]
\setlength{\epsfxsize=0.75\textwidth}
\setlength{\epsfysize=0.4\textheight}
\centerline{\epsffile{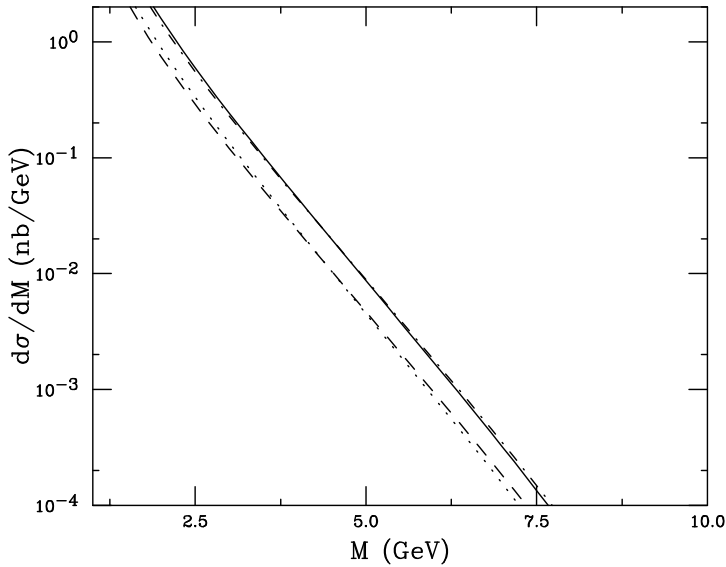}}
\caption{The Drell-Yan mass distribution in Pb+Pb collisions at 17.3 GeV 
evaluated at $Q = m$.   The solid curve is $\sigma_{\rm NLO(2)}$, 
the dashed curve is $\sigma_{\rm LO(1)}$, the dot-dashed curve is 
$\sigma_{\rm NLO(1)}$ and the dotted curve is $\sigma_{\rm LO(2)}$.  
}
\label{dymdist}
\end{figure}

Since we are interested in the makeup of the dilepton continuum in heavy
ion collisions at the SPS, the values of $\sqrt{S}$ and $M$ relevant for our
discussion are both considerably lower than in the previous section. 
We first present the Drell-Yan mass
distributions in Fig.~\ref{dymdist}.  
At $\sqrt{S} = 17.3$ GeV, the $\gamma^*
- Z$ and $Z^0$ contributions to the Drell-Yan cross section, though included, 
are negligible.
Note also that the slope of $\sigma_{\rm LO(1)}$ with $M$ is different than 
$\sigma_{\rm LO(2)}$.  In particular, $\sigma_{\rm LO(1)}$ 
is similar to and larger than 
$\sigma_{\rm LO(2)}$ for $M > 4$ GeV, where NA50 fits their data to the LO 
calculation, while, for $M < m_{\psi'}$, $\sigma_{\rm LO(2)}$ is larger than 
$\sigma_{\rm LO(1)}$.  Thus,
choosing a NLO PDF to fit an experimental $K$ factor to a LO calculation
could lead to some differences in the calculated Drell-Yan cross section
below the $J/\psi$ mass region.

\begin{figure}[htb]
\setlength{\epsfxsize=0.75\textwidth}
\setlength{\epsfysize=0.4\textheight}
\centerline{\epsffile{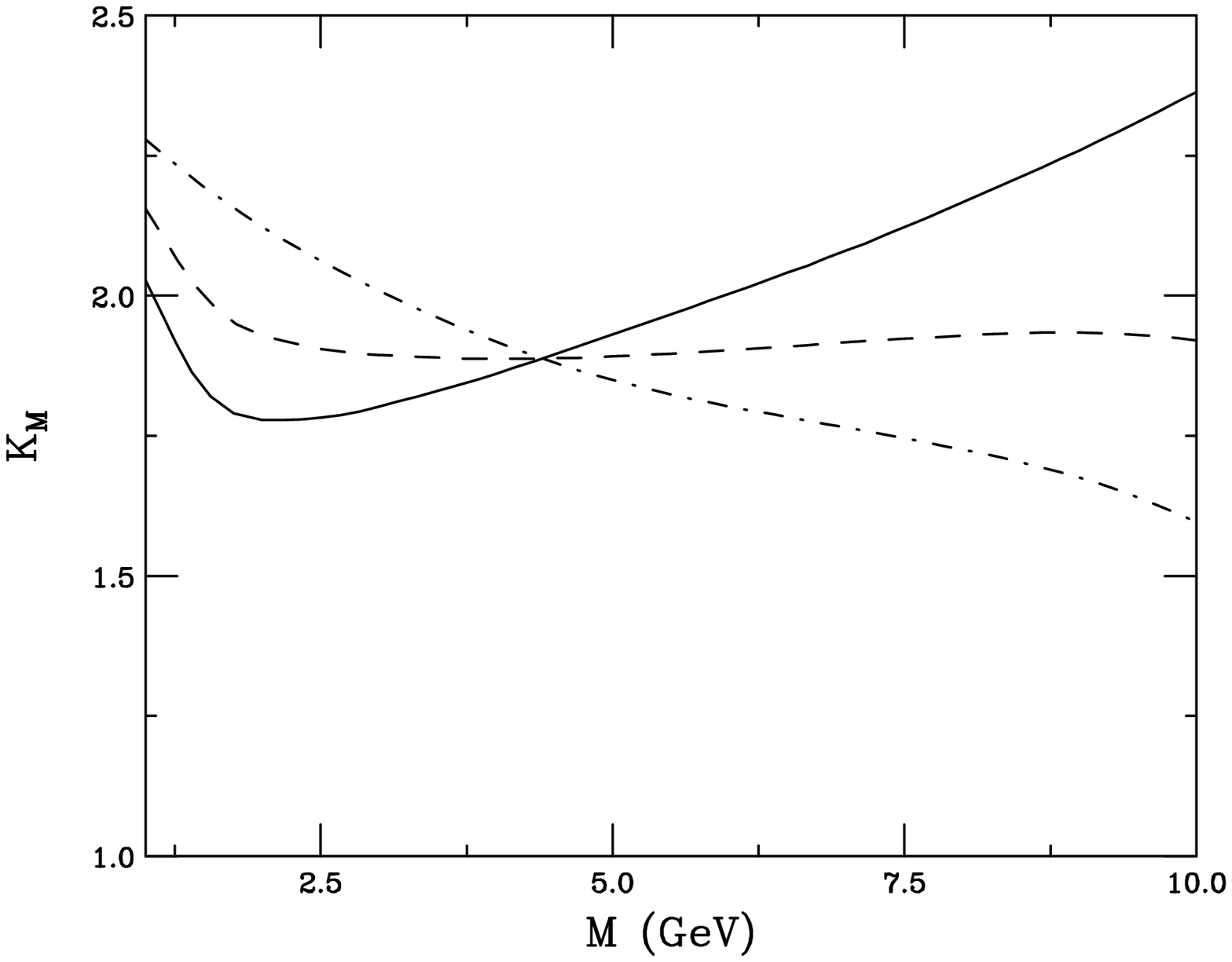}}
\caption{The three $K$ factors for Drell-Yan production at the SPS as a
function of mass.  The solid curve is
$K_{\rm th, \, 0}^{(1)}$, the dashed is $K_{\rm th, \, 1}^{(1)}$, and the 
dot-dashed is $K_{\rm th, \, 2}^{(1)}$.
}
\label{dymkfac}
\end{figure}

The calculated $K$ factors, shown in Fig.~\ref{dymkfac} as a function of $M$,
demonstrate different dependencies on $M$.  
At this lower energy, all $K$ factors are larger than for $W^+$ production in
the previous section.  Some of this difference is because $Q^2 = M^2 \ll m_W^2$
so that $\alpha_s$ is larger for low mass Drell-Yan production.
The standard $K$ factor,
$K_{\rm th, \, 0}^{(1)}$, increases from $\sim~1.75$ to 2.3 between 1 and 10
GeV while $K_{\rm th, \, 1}^{(1)}$ is nearly constant at $\sim~1.95$ and 
$K_{\rm th, \, 2}^{(1)}$ decreases over the entire mass interval.  None of 
these values agree with the constant experimental $K$ factor of 1.7 
\cite{na50} integrated over $M$ obtained by NA50
with the MRSA \cite{mrsa} set.  Since the Drell-Yan data have
relatively low statistics, a precise determination of the Drell-Yan cross
section is difficult for $M > 4$ GeV, especially when the Pb+Pb data are 
divided into centrality bins.  Thus, any measure of changes in the slope of the
mass distribution in Fig.~\ref{dymdist} is unlikely in nuclear collisions at 
this energy.  However, as shown in Figs.~\ref{dymdist} and \ref{dymkfac}, the 
relative slopes and thus
the theoretical $K$ factors do change.  A constant $K$
factor cannot account for this.  The ratios of the $J/\psi$ to Drell-Yan
cross sections used to represent the suppression pattern as a function of
$E_T$, whether in the measured Drell-Yan or in the minimum bias analyses, both
depend on extrapolation of $\sigma_{\rm LO(2)}$ multiplied by the 
experimental $K$ factor.

The results for $\sigma_{\rm LO(2)}$ and $\sigma_{\rm NLO(2)}$ have also
been calculated with the MRSA low $Q^2$ PDF \cite{mrsa}
used by NA50 \cite{na50}.  Since there is no corresponding LO PDF for the MRSA
set, we only compare $K_{\rm th,\, 0}^{(1)}$.  
We find that $\sigma_{\rm NLO(2)}$ 
with MRSA is smaller 
than that with MRST for all masses, as much as 23\% smaller at $M=10$ GeV.
On the other hand, $\sigma_{\rm LO(2)}$ is a few percent larger with MRSA for 
$M<6$ GeV, dropping below the MRST cross section at higher masses by up to
16\%. These differences are relatively small but, taken together, reduce 
$K_{\rm th, \, 0}^{(1)}$ for MRSA to $\sim 1.8$ at 4 GeV, closer to the
experimental $K$ factor of NA50.

\begin{figure}[htb]
\setlength{\epsfxsize=0.75\textwidth}
\setlength{\epsfysize=0.4\textheight}
\centerline{\epsffile{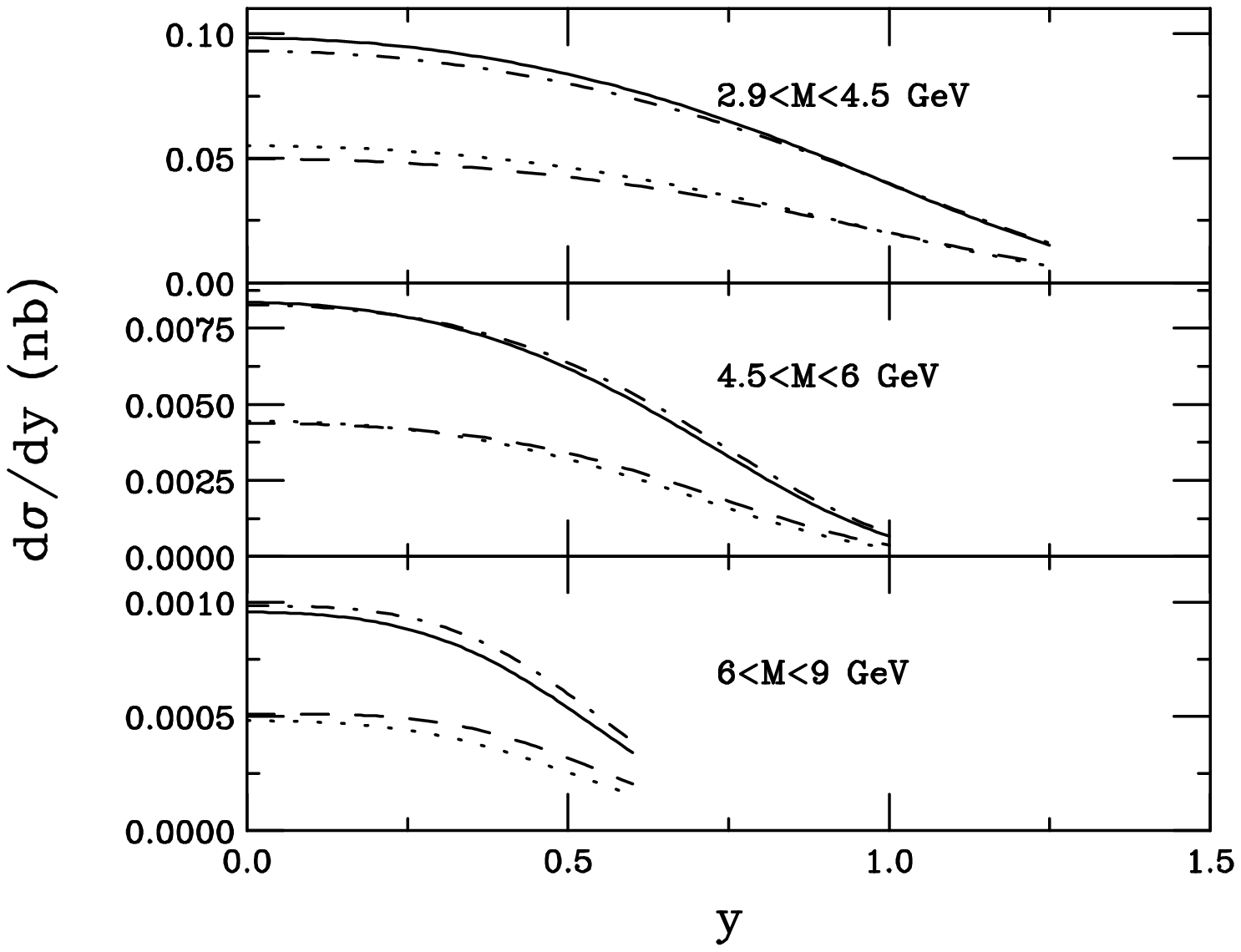}}
\caption{The Drell-Yan rapidity distribution in Pb+Pb collisions at 17.3 GeV 
evaluated at $Q = m$.   The solid curves are $\sigma_{\rm NLO(2)}$, 
the dashed curves $\sigma_{\rm LO(1)}$, the dot-dashed curves 
$\sigma_{\rm NLO(1)}$ and the dotted curves $\sigma_{\rm LO(2)}$.  
}
\label{dyydist}
\end{figure}

\begin{figure}[htb]
\setlength{\epsfxsize=0.75\textwidth}
\setlength{\epsfysize=0.4\textheight}
\centerline{\epsffile{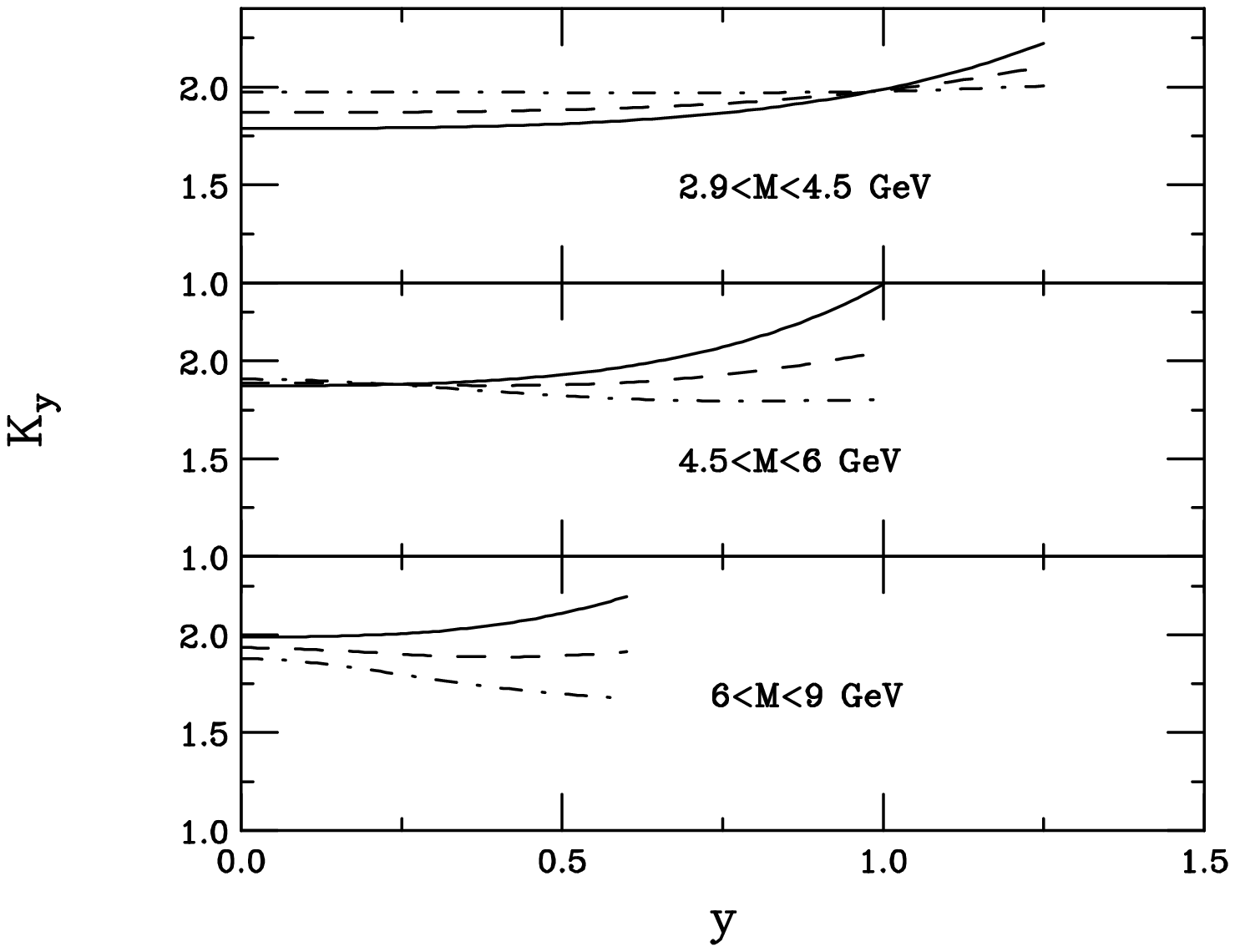}}
\caption{The three $K$ factors for Drell-Yan production at the SPS as a
function of rapidity.  The solid curves are
$K_{\rm th, \, 0}^{(1)}$, the dashed are $K_{\rm th, \, 1}^{(1)}$, and the 
dot-dashed are $K_{\rm th, \, 2}^{(1)}$.
}
\label{dyykfac}
\end{figure}

The Drell-Yan rapidity distributions and $K$ factors are shown in
Figs.~\ref{dyydist} and \ref{dyykfac} respectively.  
We have separated the
results into three mass bins: $2.9 < M<4.5$, the region in which NA50 compares
the Drell-Yan and $J/\psi$ data; $4.5 < M<6$ GeV; and $6<M<9$ GeV, above the
$J/\psi$ region.  Of the latter two bins, the lower mass bin should correspond
to the best statistics for NA50.  Unfortunately, the NA50 Drell-Yan data are
not binned in rapidity and are only given in the interval $0<y<1$.
Note that the rapidity range becomes more restrictive as the mass increases.
Only the lower two bins reach $y=1$. 

The trends of the cross
sections as a function of mass in Fig.~\ref{dymdist} are shown more clearly 
in Fig.~\ref{dyydist}.  
In the lower mass bin, $\sigma_{\rm NLO(2)} > \sigma_{\rm NLO(1)}$ and 
$\sigma_{\rm LO(2)} > \sigma_{\rm LO(1)}$ for $y<0.75$.  
In the intermediate mass bin, the two ways of
calculating the LO and NLO cross sections coincide up to $y \approx 0.5$ where
the results calculated with $\sigma_{\rm LO(1)}$ become larger.  
Finally, in the large
mass bin, $\sigma_{\rm NLO(1)} > \sigma_{\rm NLO(2)}$ and $\sigma_{\rm LO(1)} >
\sigma_{\rm LO(2)}$ for the entire rapidity range.  The $K$ factors
as a function of rapidity reflect the cross section pattern.  Again the $K$
factors are all greater than 1.7 and, except for $K_{\rm th, \, 2}^{(1)}$, are
not constant but tend to grow with rapidity, especially 
$K_{\rm th, \, 0}^{(1)}$.  

Interestingly, when the cross sections are
calculated with MRSA, the corresponding $K_{\rm th, \, 0}^{(1)}$ are
always somewhat smaller than those with MRST and $K_{\rm th, \,
0}^{(1)} \sim 1.7$ near $y \approx 0$ in the lowest mass bin.  The agreement
with NA50's experimental $K$ factor is more
likely a fortuitous accident than a precise fit.  Given the rather limited
Drell-Yan statistics of NA50 for $M>4$ GeV, it is doubtful that a better
measure of the experimental $K$ factor could have been obtained.  It would
be worthwhile, however, to calculate the full NLO cross section, either 
$\sigma_{\rm NLO(1)}$ or $\sigma_{\rm NLO(2)}$ depending on the chosen PDF, and
obtain the experimental $K$ factor at NLO.  Our results suggest that it would
be significantly smaller but still above unity, $\sim 1.2$.

\section{Heavy Quark Production in $pp$ Interactions}
\label{heavyq}

We now turn to evaluations of the heavy quark cross sections and their $K$
factors.  We concentrate
only on the bare distributions rather than introduce empirical fragmentation
functions which, for heavy quarks, remain at a rather primitive level
\cite{Pete}. We can expect the biggest variations in the $K$ factors 
here because
the LO cross section is already proportional to $\alpha_s^2$.  
The order at which $\alpha_s$ is calculated is important because the lower
$Q^2$, proportional to $m_T^2 = m_Q^2 + p_T^2$ in distributions, 
means $\alpha_s$ is much 
larger for charm and bottom quarks
than for gauge bosons.  Differences in $\Lambda_{\rm QCD}$ between LO and NLO
PDF fits result in similar values of $\alpha_s$ at one and two loops when the
LO and NLO values of $\Lambda_{\rm QCD}$ respectively are used.
However, large discrepancies may result if the LO value of $\Lambda_{\rm QCD}$
is used in a two-loop evaluation of $\alpha_s$.  Thus, one must be careful
to use PDFs and $\alpha_s$ evaluations that are compatible.  

Heavy quarks are produced at LO by $gg$ fusion and
$q \overline q$ annihilation while at NLO $g(q + \overline q)$ 
scattering is also included.  To any order, the partonic 
cross section may be expressed in terms of dimensionless scaling functions
$f^{(k,l)}_{ij}$ that depend only on the variable $\eta = s/4m_Q^2 - 1$ 
\cite{KLMV},
\begin{eqnarray}
\label{scalingfunctions}
\sigma_{ij}(s,m_Q^2,Q^2) = \frac{\alpha^2_s(Q^2)}{m_Q^2}
\sum\limits_{k=0}^{\infty} \,\, \left( 4 \pi \alpha_s(Q^2) \right)^k
\sum\limits_{l=0}^k \,\, f^{(k,l)}_{ij}(\eta) \,\,
\ln^l\left(\frac{Q^2}{m_Q^2}\right) \, , 
\end{eqnarray} 
where $s$ is the partonic center of mass energy squared, 
$m_Q$ is the heavy quark mass and
$Q^2$ is the scale.
The cross section is calculated as an expansion in powers of $\alpha_s$
with $k=0$ corresponding to $\sigma_{\rm LO} = \sigma(\alpha_s^2)$.  
The first correction, $k=1$, corresponds to $\sigma(\alpha_s^3)$. 
Note that no distinction is made
between LO and
higher evaluations of $\alpha_s$ in Eq.~(\ref{scalingfunctions}).  The scale
$Q^2$ is generally assumed to be the same for both the renormalization, 
$Q_R^2$, and factorization, $Q_F^2$, scales
since all PDF analyses make the assumption that $Q_R^2 = Q_F^2 = Q^2$.  
It is only when $k \geq 1$ that
the dependence on $Q_R^2$ can be distinguished
since when $k=1$
and $l=1$, the logarithm $\ln(Q^2/m_Q^2)$ 
appears.  The dependence on $Q_F^2$ appears already at LO in the argument of
$\alpha_s$ in the partonic cross section and in the parton densities.  
The NNLO corrections to next-to-next-to-leading logarithm with $k=2$
have been calculated near threshold \cite{KLMV}.
The complete calculation only exists to NLO.
The total hadronic cross section is obtained by convoluting the total partonic
cross section with the PDFs,
\begin{eqnarray}
\label{totalhadroncrs}
\sigma_{pp}(S,m_Q^2) = \sum_{i,j = q,{\overline q},g} 
\int_{\frac{4m_Q^2}{S}}^{1} d\tau \int_{\tau}^{1} \frac{dx_1}{x_1} \, 
f_i^p(x_1,Q^2) f_j^p\bigg(\frac{\tau}{x_1},Q^2 \bigg) \, 
\sigma_{ij}(\tau S,m_Q^2,Q^2)\,  
\end{eqnarray}
where the sums $i$ and $j$ are over all massless partons.

\begin{figure}[htb]
\setlength{\epsfxsize=0.95\textwidth}
\setlength{\epsfysize=0.25\textheight}
\centerline{\epsffile{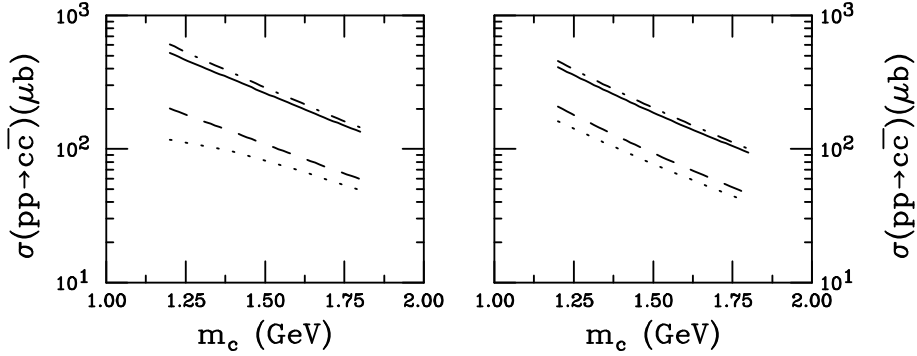}}
\caption[]{The $c \overline c$ total cross sections as a function of charm 
quark mass in $pp$ collisions at 200 GeV.  The lefthand plot shows the results
for $Q^2 = m_c^2$ while the righthand plot is calculated with $Q^2 = 4m_c^2$.
The solid curves are $\sigma_{\rm NLO(2)}$, the dashed curves $\sigma_{\rm
LO(1)}$, the dot-dashed curves
$\sigma_{\rm NLO(1)}$ and the dotted curves 
$\sigma_{\rm LO(2)}$.  
}
\label{sigmc}
\end{figure}

\begin{figure}[htb]
\setlength{\epsfxsize=0.95\textwidth}
\setlength{\epsfysize=0.5\textheight}
\centerline{\epsffile{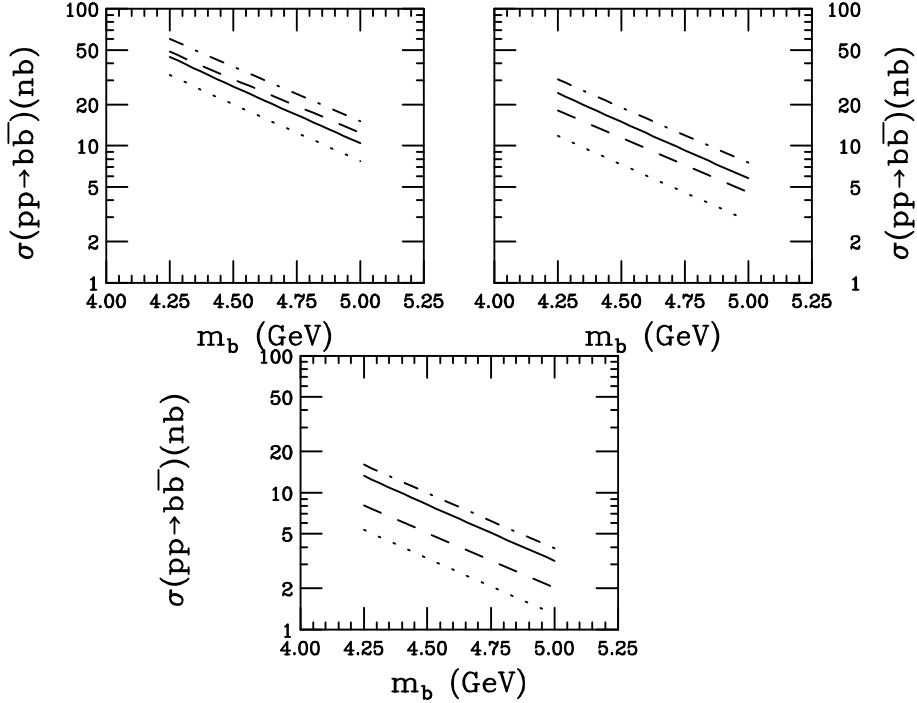}}
\caption[]{The $b \overline b$ total cross sections as a function of bottom 
quark mass in $pp$ collisions at 41.6 GeV. Clockwise from the upper left, the
results are calculated with $Q^2 = m_b^2/4$, $m_b^2$, and $4m_b^2$ 
respectively.
The solid curves are $\sigma_{\rm NLO(2)}$, the dashed curves $\sigma_{\rm
LO(1)}$, the dot-dashed curves 
$\sigma_{\rm NLO(1)}$ and the dotted curves 
$\sigma_{\rm LO(2)}$.  
}
\label{sigmb}
\end{figure}

To illustrate the dependence of the total $Q \overline Q$ cross sections
upon $m_Q$ and $Q^2$, we show the $c \overline c$
and $b \overline b$ cross sections as a function of mass and scale in
Figs.~\ref{sigmc} and \ref{sigmb}.  
The charm mass is varied between 1.2 GeV
and 1.8 GeV in $\sqrt{S} = 200$ GeV $pp$ collisions in Fig.~\ref{sigmc}.  
We choose this energy
because it is close to that of the PHENIX charm measurement at RHIC
\cite{PXcharm}, $\sqrt{S} = 130$ GeV.  A more complete measurement should be 
coming from the 200 GeV data taken in the latest run.  The bottom mass is
varied between 4.25 GeV and 5 GeV for $pp$ collisions at $\sqrt{S}=41.6$ 
GeV, the energy
of the HERA-B experiment which has recently presented a measurement of the
$b \overline b$ total cross section \cite{herab}.  Unlike charm production at
RHIC, bottom production at this energy is in the near-threshold region where
resummation techniques can be applied \cite{KLMV}.

The $c \overline c$ cross section as a function of mass in Fig.~\ref{sigmc} 
is calculated for two different values of the scale, $Q^2 =
m_c^2$ (left-hand side) and $4m_c^2$ (right-hand side).  Note that only scales
greater than $m_c^2$ 
are shown because $Q^2 = m_c^2/4 < 1$ GeV$^2$ for all charm masses 
considered, below the minimum scale of most PDFs.  Thus results for scales
lower than $m_c^2$ would not be particularly meaningful.  For this particular
energy, the scale dependence is not strong, 10-20\% at LO for $m_c = 1.5$ GeV,
and 40\% at NLO.  The stronger scale dependence at NLO is unusual and shows
that charm could be difficult to treat at higher orders because of its
relatively low mass \cite{smithrv}.  
The mass dependence is stronger for the NLO calculations
than the LO and is also stronger for the larger scale.  Note that in both cases
displayed in Fig.~\ref{sigmc} $\sigma_{\rm NLO(2)} < \sigma_{\rm NLO(1)}$ 
because $\sigma_{\rm LO(1)} > \sigma_{\rm LO(2)}$.  The difference is small
since  $K_{\rm th}^{(1)}>2$ for all three definitions of the $K$ factor.
Thus $\sigma(\alpha_s^3)$ alone is significantly larger than either calculation
of the LO cross section.  The most important sources of the difference in the
LO cross sections are the larger LO gluon density and the dominance of the $gg$
production channel.  The charm $K$ factors tend to be larger for lower masses
and smaller scales.

We have shown the results for one specific energy here.  However, the scale
dependence changes with energy, as already noted in Section~\ref{wp}.  
The growth of the $c \overline c$ total cross sections with energy is slower
for $Q^2 = m_c^2$.  At lower energies, $\sigma(m_c^2) > \sigma(4m_c^2)$.  As
the energy increases, $\sigma(4m_c^2) > \sigma(m_c^2)$.  The value of
$\sqrt{S}$ at which this cross over occurs changes with $m_c$, $\sqrt{S}
\approx 100$ GeV for $m_c = 1.2$ GeV and $\sqrt{S} \sim 3$ TeV for $m_c = 1.8$
GeV. The change in energy dependence of the cross section with scale is 
due to the evolution of
the PDFs at large $Q^2$ and low $x$, similar to what was observed for $W^+$
production in Section~\ref{wp}.  

In Fig.~\ref{sigmb}, the $b \overline b$ cross section is given as a function
of mass for three different values of $Q^2$, $m_b^2/4$ (upper left), $m_b^2$ 
(upper right) and $4m_b^2$ (bottom center).  
The larger bottom quark mass now allows us
to calculate the cross section for scales lower than $m_b^2$ and still obtain
meaningful results.  In this case, the scale dependence is reduced at NLO
relative to LO.  The mass dependence is stronger at NLO, as for charm, and
increases somewhat with scale.  The heavier quark mass reduces the impact of
$\sigma(\alpha_s^3)$ on the total cross sections.  Indeed, for $Q^2 = m_b^2/4$,
$\sigma_{\rm LO(1)} > \sigma_{\rm NLO(2)}$.  This
effect is probably due to the one-loop $\alpha_s$ evaluation in
$\sigma_{\rm LO(1)}$ since the one-loop $\alpha_s$ is larger than the two-loop
$\alpha_s$, particularly at low scales.  
The bottom $K$ factors are all smaller than those for
charm, $K_{\rm th}^{(1)}< 2$ in most cases. 
Thus $b \overline b$ production, at least
at this energy, is better under control than charm.  Like charm, however,
the $K$ factors also are larger for lower masses and higher scales.  The scale
dependence of $b \overline b$ production with energy is smaller than for charm.
The bottom cross sections follow the hierarchy $\sigma(m_b^2/4) > \sigma(m_b^2)
> \sigma(4m_b^2)$ for all energies and orders.  
Near threshold the difference in cross
sections due to scale can be nearly a factor of 10 but by $\sqrt{S} = 14$ TeV,
the difference is only a few percent.  The smaller scale dependence is also an
indication that bottom production is better under control than charm.

The mass and scale parameters used in our further calculations are
determined by obtaining a `best' fit to the data without an
experimental $K$ factor.  Using the MRST PDFs to calculate 
$\sigma_{\rm NLO(2)}$, we found $m_c =
1.2$ GeV and $Q^2 = 4m_c^2$ and $m_b = 4.75$ GeV 
with $Q^2 = m_b^2$ for bottom \cite{RVhpc,RVww,RVbud}.  
Although there is a great deal of scattered data 
up to $\sqrt{S} = 63$ GeV on the $c \overline c$ total cross section, our
choice for the bottom parameters are in line with conventional wisdom rather
than data.
When other PDFs were used, the parameters favored
varied somewhat \cite{RVww,RVbud} 
but, in the following, we compare results from different PDFs
using the same values of $m_Q$ and $Q^2$.

Since the charm mass obtained from these evaluations is somewhat smaller than
the 1.5 GeV mass generally used and the NLO calculations with $m_b = 4.75$ GeV
underestimate the Tevatron $p \overline p \rightarrow b \overline b$ data
\cite{cdf}, it is reasonable to expect that higher order
corrections beyond NLO could be large.  Indeed, the HERA-B 
cross section \cite{herab} agrees with the NNLO-NNLL cross section in 
Ref.~\cite{KLMV}, suggesting that the next order correction could be nearly a
factor of two.  Thus the NNLO correction could be nearly as large as the total
NLO cross section.

Unfortunately, the NNLO-NNLL calculation is 
valid only near threshold.  The $p \overline p$ data at higher energies,
while not total cross sections, also show a large discrepancy between the
perturbative NLO result and the data, nearly a factor of three \cite{cdf}.  
This difference could be accounted for using unintegrated parton densities
\cite{unint} although these unintegrated distributions vary widely.  
Another, more mundane explanation to the $b \overline b$ discrepancy
is an incomplete understanding of the hadronization process \cite{matteo}.
If some resummation is needed in the high energy regime to account for the
Tevatron data, there remains the difficulty of connecting the regimes where 
near-threshold corrections are applicable and where high-energy, small $x\sim
m_Q/\sqrt{S}$ physics dominates.  This problem is increased for charm where, 
even at low energies, we are far away from threshold.  

Our method is perhaps the
most straightforward--using $\sigma_{\rm NLO(2)}$ 
and ignoring higher-order corrections to
fit the data.  This is not difficult for $c \overline c$ because 
the data are extensive.  However, there
are less $b \overline b$ data.  The $\pi^- p \rightarrow c \overline
c$ data tend to favor lighter charm
masses.  It is difficult to say if the same is
true for $b \overline b$.  A value of $m_b = 4.75$ GeV, which underpredicts
the Tevatron results compared to NLO cross sections \cite{cdf}, agrees 
reasonably well with the average of the $\pi^- p$ data.  
However, for the HERA-B
measurement to be compatible with a NLO evaluation, the $b$ quark mass would
have to be reduced to 4.25 GeV, a value which might be more compatible with
the Tevatron results.  Therefore, a full NNLO calculation with $m_c \sim 1.5$
GeV and $m_b \sim 4.75$ GeV might agree with the $Q \overline Q$ data without 
an experimental $K$ factor.
A more quantitative statement is not possible.

\begin{figure}[htb]
\setlength{\epsfxsize=0.95\textwidth}
\setlength{\epsfysize=0.5\textheight}
\centerline{\epsffile{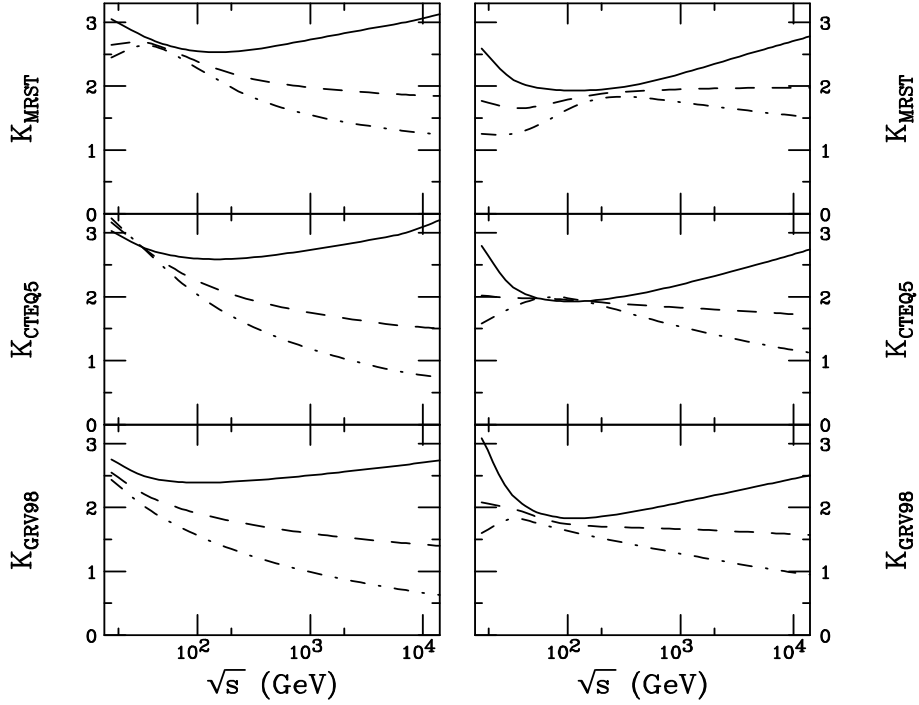}}
\caption[]{The three $K$ factors for $Q \overline Q$ production in $pp$
collisions as a  function of energy.  The $c \overline c$ results, calculated
with $m_c = 1.2$ GeV and $Q^2 = 4m_c^2$, are given
on the lefthand side, while the $b \overline b$, calculated with $Q^2 = m_b^2$
and $m_b = 4.75$ GeV, are on the righthand side.
From top to bottom the results are given for the MRST, CTEQ5, and GRV 98 PDFs.
The solid curves are
$K_{\rm th, \, 0}^{(1)}$, the dashed are $K_{\rm th, \, 1}^{(1)}$, and the 
dot-dashed are $K_{\rm th, \, 2}^{(1)}$.
}
\label{qkfacs}
\end{figure}

We now employ our inferred values of $m_Q$ and $Q^2$ to calculate the $K$
factors as functions of energy and PDF.  Figure~\ref{qkfacs} illustrates the
danger inherent in calculating, $\sigma_{\rm LO(1)}$ and
then multiplying either by $K_{\rm th, \, 0}^{(1)}$ or by an arbitrary factor
of 2:  the cross sections can be considerably overestimated.  
Indeed, $K$
factors are generally applied to avoid underestimating the cross sections.
We see that $K_{\rm th, \, 0}^{(1)}$ is generally greater than 2 at low 
energies, drops somewhat and finally increases with energy.  
This dependence of the
standard $K$ factor could be attributed to more inherent theoretical
uncertainties in low
and high energies where different resummation techniques are applicable.  There
is a somewhat stronger effect for $b \overline b$ since $2m_b/\sqrt{S} \geq
0.1$, near the threshold region, until $\sqrt{S} \sim 100$ GeV.  On the other
hand, charm production is well above threshold for all energies shown since
$2m_c/\sqrt{S} \leq 0.1$ for $\sqrt{S} > 20$ GeV.  Note that $K_{\rm th, \,
1}^{(1)}$ and $K_{\rm th, \, 2}^{(1)}$ both decrease with energy with a larger
decrease for $c \overline c$ (lefthand plots) than $b \overline b$ (righthand
plots).  This decrease can clearly be attributed to decreasing $x \sim
\sqrt{Q^2/S}$ with increasing $\sqrt{S}$.  The $x$ range of charm production 
is from $x \sim 0.1$ at 15 GeV to $10^{-4}$ at 14 TeV.  The LO gluon PDF is
larger than the NLO gluon PDF, particularly at low scales, so that
$\sigma_{\rm LO(1)}$ increases faster than $\sigma_{\rm NLO(2)}$ 
and $K_{\rm th, \, 2}^{(1)}$ decreases to near or below
unity at high energies.  The values of $x$ and $Q^2$ are larger for 
$b \overline b$ production so that the decrease in $K_{\rm th, \, 2}^{(1)}$
with energy is slower.  $K_{\rm th, \, 1}^{(1)}$ does not
decrease as quickly as $K_{\rm th, \, 2}^{(1)}$, and is almost independent
at high energies, but is still smaller than
$K_{\rm th, \, 0}^{(1)}$.  Thus at high energies, $\sigma_{\rm LO(1)} \sim
\sigma_{\rm NLO(2)}$ in some cases and including a
$K$ factor on $\sigma_{\rm LO(1)}$ would overestimate the cross section by a
factor of $2-3$.  The same trends are observed for results 
with the CTEQ5 and GRV 98 PDFs.

\begin{figure}[htb]
\setlength{\epsfxsize=0.95\textwidth}
\setlength{\epsfysize=0.4\textheight}
\centerline{\epsffile{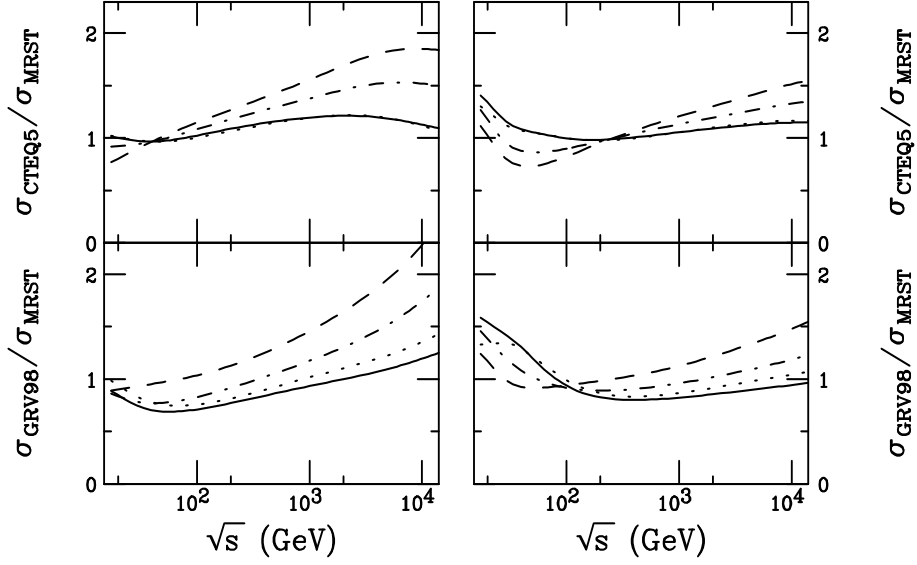}}
\caption[]{Ratios of the total $Q \overline Q$ cross sections calculated with
the CTEQ5 (top) and GRV 98 (bottom) PDFs to those calculated with the MRST
PDFs as a function of energy. The $c \overline c$ results, calculated
with $m_c = 1.2$ GeV and $Q^2 = 4m_c^2$, are given
on the lefthand side, while the $b \overline b$, calculated with $Q^2 = m_b^2$
and $m_b = 4.75$ GeV, are on the righthand side.
The solid ratios are $\sigma_{\rm NLO(2)}$, 
the dashed ratios $\sigma_{\rm LO(1)}$, the 
dot-dashed ratios 
$\sigma_{\rm NLO(1)}$ and the dotted ratios 
$\sigma_{\rm LO(2)}$.  
}
\label{qpdfrat}
\end{figure}

The PDF dependence is illustrated in Fig.~\ref{qpdfrat} with the ratios of the
CTEQ5 and GRV 98 LO and NLO cross sections to the MRST cross
sections.  
The ratios depend most strongly on the relative gluon PDFs.  The
CTEQ5 and MRST NLO gluon distributions are very similar so that the ratios of
$\sigma_{\rm LO(2)}$ and $\sigma_{\rm NLO(2)}$ are
nearly unity without much variation over the entire energy range.  There are
slight differences for $b \overline b$ at low energies due to the larger $q
\overline q$ contribution near threshold.  The GRV 98 NLO gluon distribution is
less similar to the MRST gluon, leading to a larger variation of the ratios
with energy.  The ratios involving the LO gluon PDFs are larger and more energy
dependent because the LO gluon distribution is not as well constrained.  The
biggest differences are in the ratios of $\sigma_{\rm LO(1)}$ since 
the large $gg$ contribution to $\sigma(\alpha_s^3)$ reduces the 
effect of the LO gluon distribution in the $\sigma_{\rm NLO(1)}$ ratios.

\begin{figure}[htb]
\setlength{\epsfxsize=0.95\textwidth}
\setlength{\epsfysize=0.5\textheight}
\centerline{\epsffile{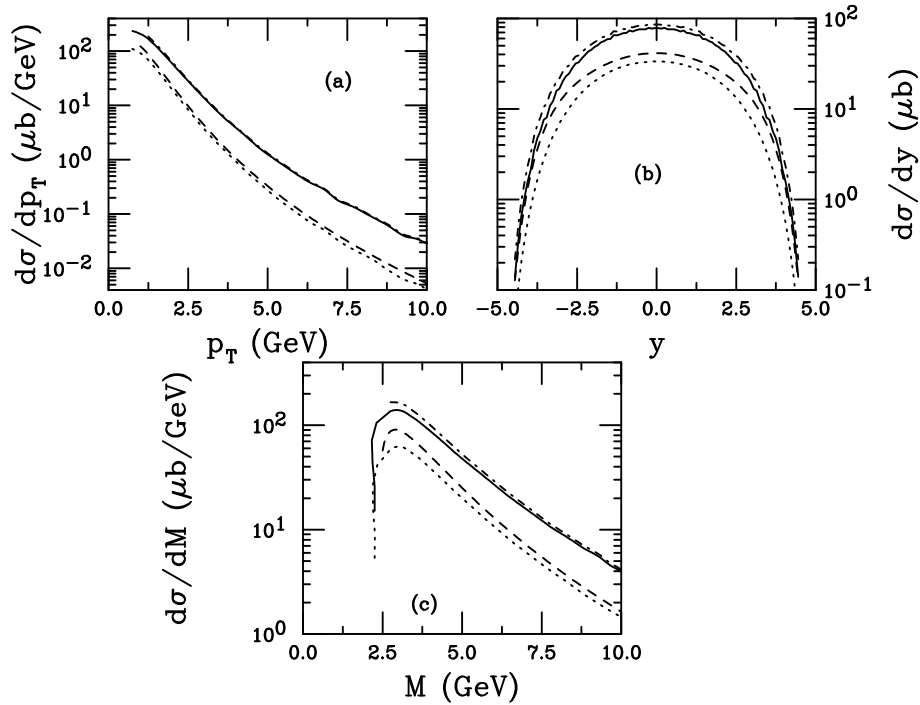}}
\caption[]{The charm quark $p_T$ (a), rapidity (b), and $c \overline c$ pair
mass (c) for $pp$ collisions at 200 GeV.
The solid curve is $\sigma_{\rm NLO(2)}$, the dashed curve is $\sigma_{\rm
LO(1)}$, the dot-dashed curve is 
$\sigma_{\rm NLO(1)}$ and the dotted curve is 
$\sigma_{\rm LO(2)}$.  
}
\label{crdist}
\end{figure}

\begin{figure}[htb]
\setlength{\epsfxsize=0.95\textwidth}
\setlength{\epsfysize=0.5\textheight}
\centerline{\epsffile{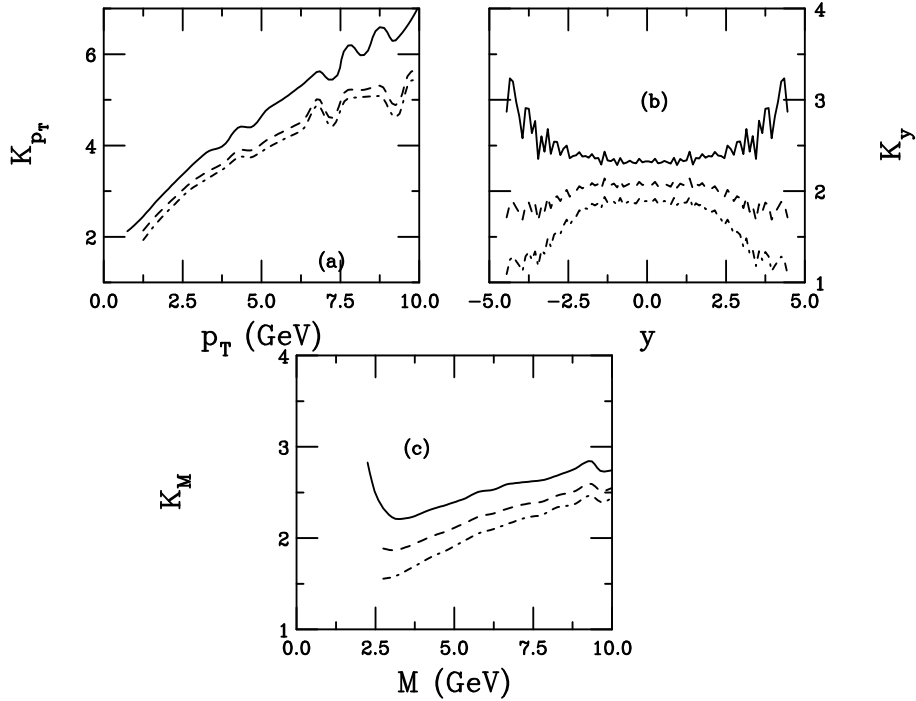}}
\caption{The three $K$ factors  as a function of charm quark $p_T$ (a), 
rapidity (b), and $c \overline c$ pair
mass (c) for $pp$ collisions at 200 GeV.  The solid curve is
$K_{\rm th, \, 0}^{(1)}$, the dashed is $K_{\rm th, \, 1}^{(1)}$, and the 
dot-dashed is $K_{\rm th, \, 2}^{(1)}$.
}
\label{crkfac}
\end{figure}

\begin{figure}[htb]
\setlength{\epsfxsize=0.95\textwidth}
\setlength{\epsfysize=0.5\textheight}
\centerline{\epsffile{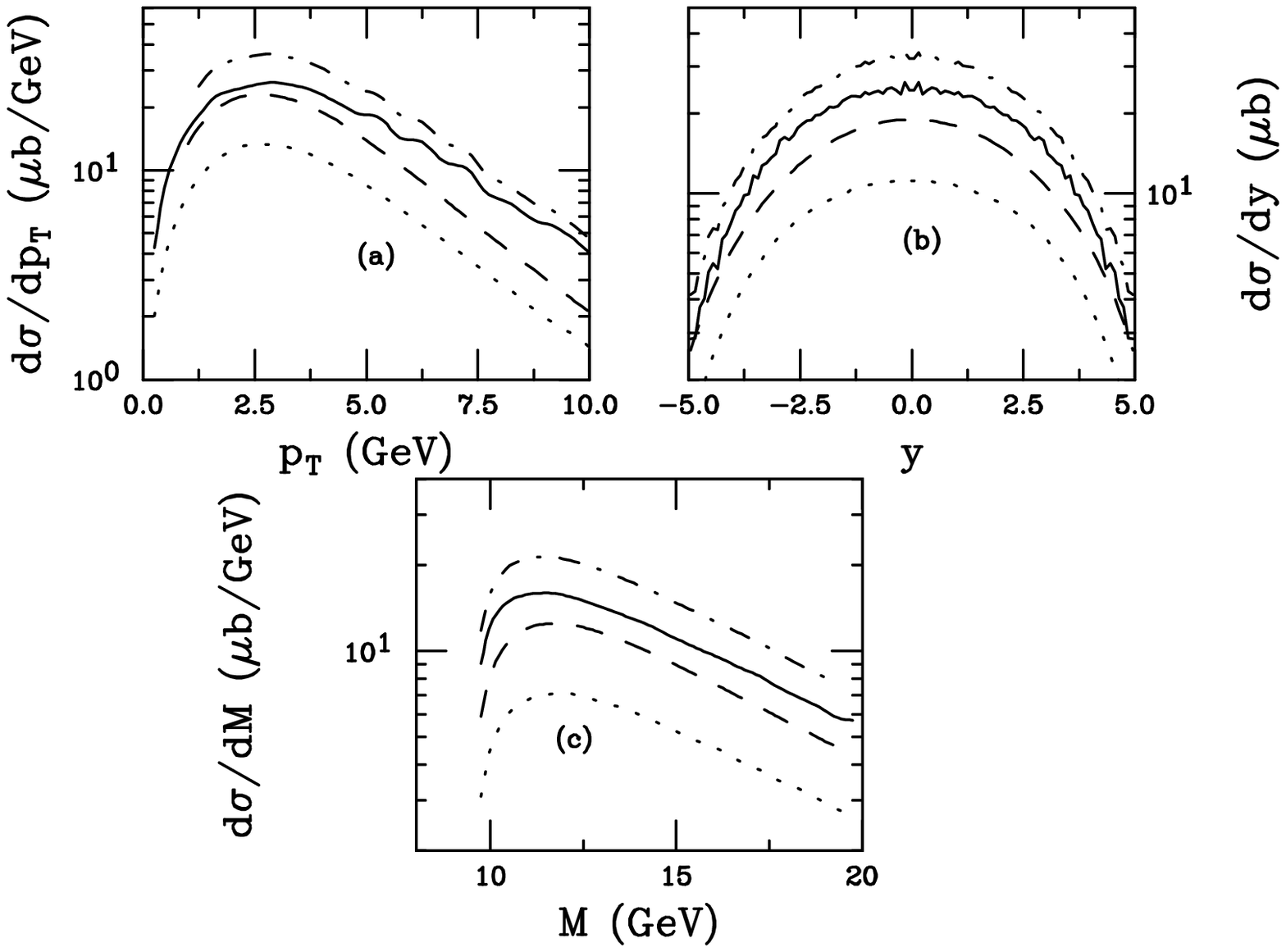}}
\caption[]{The bottom quark $p_T$ (a), rapidity (b), and $b \overline b$ pair
mass (c) for $pp$ collisions at 5.5 TeV.
The solid curve is $\sigma_{\rm NLO(2)}$, the dashed curve is $\sigma_{\rm
LO(1)}$, the dot-dashed curve is $\sigma_{\rm NLO(1)}$ and the 
dotted curve is $\sigma_{\rm LO(2)}$.  
}
\label{bldist}
\end{figure}

\begin{figure}[htb]
\setlength{\epsfxsize=0.95\textwidth}
\setlength{\epsfysize=0.5\textheight}
\centerline{\epsffile{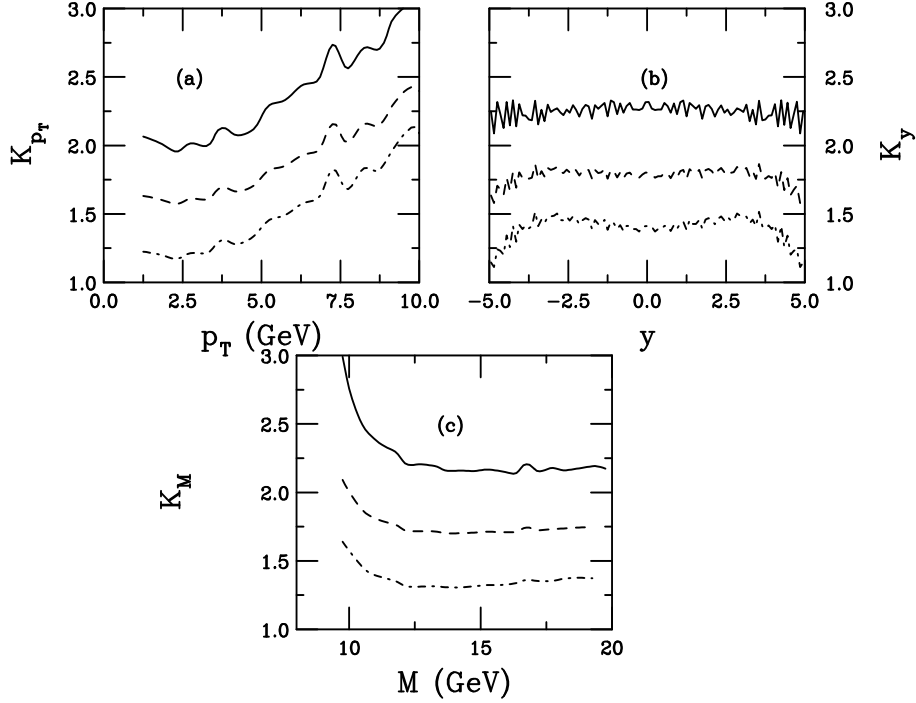}}
\caption{The three $K$ factors  as a function of bottom quark $p_T$ (a), 
rapidity (b), and $b \overline b$ pair
mass (c) for $pp$ collisions at 5.5 TeV.  The solid curve is
$K_{\rm th, \, 0}^{(1)}$, the dashed is $K_{\rm th, \, 1}^{(1)}$, and the 
dot-dashed is $K_{\rm th, \, 2}^{(1)}$.
}
\label{blkfac}
\end{figure}

To complete our discussion of heavy quarks, we now turn to the charm and bottom
distributions and their corresponding $K$ factors, shown in
Figs.~\ref{crdist}-\ref{blkfac}. 
The necessity of NLO evaluations of $Q \overline Q$ production is clear if one
wants to study pair production and correlations.  The $p_T$ of the $Q \overline
Q$ pair is zero at LO and only becomes finite at NLO.  LO calculations 
are thus only useful for obtaining the single quark distributions in $p_T$,
rapidity and Feynman $x$, $x_F$, and for the $Q \overline Q$ pair, the $y$, 
$x_F$ and invariant mass, $M$,
distributions.  Therefore the quantities for which we calculate the $K$ factor,
besides the total cross sections,
are the single quark $p_T$ and rapidity distributions and the pair
mass distribution.  We will evaluate the differential
$K$ factors for charm distributions
at RHIC, $\sqrt{S} = 200$ GeV, and bottom at the LHC, $\sqrt{S} = 5.5$ TeV,
both appropriate energies for ion colliders.
We calculate $\sigma_{\rm LO(1)}$
analytically, resulting in smooth distributions.  However,
$\sigma_{\rm LO(2)}$ and $\sigma(\alpha_s^3)$ are calculated with the NLO
Monte Carlo code
described in Ref.~\cite{MNR} which uses a two-loop evaluation of $\alpha_s$ as
a default.  The finite statistics of the Monte
Carlo result in some fluctuations in the distributions, particularly apparent
in the $K$ factors.

The charm calculations at $\sqrt{S} = 200$ GeV for RHIC correspond to
the total cross sections shown in Fig.~\ref{sigmc} with $m_c = 1.2$ GeV and 
$Q^2 = 4m_c^2$.  
It is clear that the LO $p_T$ distributions in Fig.~\ref{crdist}
have a steeper slope than the NLO distributions.  The charm rapidity and pair
mass distributions, on the other hand, are more similar.  The $K$
factors increase rapidly with $p_T$ in Fig.~\ref{crkfac} even though we have
used $m_T^2$ as a more appropriate
scale for the distributions \cite{RVoldk}.  This increase can be expected since
$\sigma(\alpha_s^3)$
is inherently at higher $p_T$ due to contributions from
$gg \rightarrow gg^* \rightarrow gQ \overline Q$ where the $Q \overline Q$
pair is opposite a gluon jet.  Contributions such as these begin to be
important when $p_T > m_Q$, as is obvious in Fig.~\ref{crkfac}.  The increase
in the $K$ factor with $p_T$ is also only important when $x$ is relatively
small---the $K$ factors for $b \overline b$ production at RHIC increase very
slowly with $p_T$ over the same interval.  
The $p_T$ dependence of the $K$ factors clearly should be
taken into account when scaling up a LO calculation.
Conversely, the $K$ factors are nearly constant with rapidity
and pair mass which are $p_T$-integrated and are weighted by the low $p_T$ part
of the cross section.  The same trend with rapidity seen in 
Figs.~\ref{wpkfac} and \ref{dyykfac} is also
observed here:  $K_{\rm th, \, 0}^{(1)}$ increases slowly with
rapidity while $K_{\rm th, \, 2}^{(1)}$ decreases.  The exact behavior at large
rapidity is difficult to determine because the Monte Carlo fluctuations are
largest near the edges of phase space. 

A careful examination of the $K$ factors
may reveal some slight difference in magnitude between Figs.~\ref{crkfac} and
\ref{qkfacs} at the same energy.  This shift is due to the somewhat different 
scales used to calculate total cross sections and distributions.  For total
cross sections, the only relevant scale is $Q^2 \propto m_Q^2$ 
in the $\ln(Q^2/m_Q^2)$ term in
Eq.~(\ref{scalingfunctions}).  However, when calculating distributions,
$p_T$-dependent logarithms also enter which need to be controlled by the $m_T$
scale.  Shifting the scale from $\propto m_Q^2$ to $\propto m_T^2$ 
has the effect
of modifying the total cross section obtained by integrating the distributions.
This shift in scale is manifested by a change in the magnitude of the
$K$ factors as well.

The results for $b \overline b$ production in $\sqrt{S} = 5.5$ 
TeV $pp$ collisions
at the LHC are shown in Figs.~\ref{bldist} and \ref{blkfac}.  The $K$ factors
clearly begin to increase for $p_T > 5$ GeV where $p_T > m_b$.  The
increase with $p_T$ is slower than for charm production.  The $K$ factors are
essentially constant over the rapidity range shown since the edge of phase
space has not yet been reached at $y = 5$.  Also, away from 
threshold, $2m_b$,
the $K$ factors are constant as a function of mass.  Note that
$K_{\rm th, \, 0}^{(1)}$ exhibits the strongest threshold effect near $M\sim
10$ GeV.

\section{Summary}

We have discussed three ways of defining the theoretical $K$ factor to NLO.  
The first, $K_{\rm th, \, 0}^{(1)}$, is most appropriate for determining the
theoretical uncertainties between cross sections of different orders because
it allows the most straightforward determination of the next-order effects.
In cases where the use of the LO cross section is necessary for speed of 
calculation, such as in event generators, it is most appropriate to make a
full LO evaluation and then multiply by either $K_{\rm th, \, 1}^{(1)}$ or
$K_{\rm th, \, 2}^{(1)}$. 

We have shown that the theoretical $K$ factor 
is only approximately constant away
from threshold and away from the edges of phase space.  Thus for high energy
$W^+$ production, $K$ is only independent of rapidity for $y<2$.  The Drell-Yan
$K$ factor is not constant with mass, increasing at low masses and also 
changing at high masses, increasing or decreasing depending on which definition
is used.  The rapidity range over which $K$ is constant for Drell-Yan 
production at the SPS decreases with increasing mass as the edge of phase space
is approached at lower rapidities.  The $K$ factors for heavy quarks are most
strongly dependent on $p_T$ since the NLO corrections are large for $p_T \geq
m_Q$.  Similarly, a strong dependence of 
$K_{{\rm th}, \, 0}^{(1)}$ on $p_T$ in inclusive jet
production was found in Ref.~\cite{failevai}.

Clearly a constant $K$ factor is inappropriate for all kinematic variables.
To best utilize knowledge of the next-to-leading and higher order corrections,
the differential $K$ factor should be determined as a 
function of the kinematic variables
of interest, such as $p_T$, $y$ or $M$. Whether $K_{\rm th, \, 1}^{(1)}$ or
$K_{\rm th, \, 2}^{(1)}$ is used with $\sigma_{\rm LO(1)}$
is somewhat a matter of taste.  However, it
should perhaps be kept in mind that the PDFs are evaluated to NLO using
$\sigma_{\rm NLO(2)}$.
Thus $K_{\rm th, \, 2}^{(1)}$ is perhaps the most relevant for extending full
LO calculations to NLO. Therefore there is no clearcut answer to the question
posed in the title.
Whatever definition is chosen, it is obviously
important that the determination of $K_{\rm th}$ be clearly described and 
applied consistently in a given calculation.

{\bf Acknowledgements} I thank G. Fai, P. Levai, and I. Vitev for encouragement
and helpful discussions.

\end{document}